\documentclass[a4paper,11pt,preprint]{article}
\pdfoutput=1 

\usepackage{jheppub} 

\usepackage[T1]{fontenc} 

\usepackage{graphicx}
\usepackage{amsmath}
\usepackage{amssymb}
\usepackage{array}

\def\ba{\begin{eqnarray}}
\def\ea{\end{eqnarray}}

\def\be{\begin{equation}}
\def\ee{\end{equation}}

\newcommand{\checked}[1]{}

\newcommand{\la}{\label}

\usepackage{lineno}

\title{Bottomonium suppression and elliptic flow using Heavy Quarkonium Quantum Dynamics}

\author{Ajaharul Islam}
\author{and Michael Strickland}

\affiliation{Department of Physics, Kent State University, Kent, OH 44242, United States}
\emailAdd{aislam2@kent.edu}
\emailAdd{mstrick6@kent.edu}

\abstract{
We introduce a framework called Heavy Quarkonium Quantum Dynamics (HQQD) which can be used to compute the dynamical suppression of heavy quarkonia propagating in the quark-gluon plasma using real-time in-medium quantum evolution.  Using HQQD we compute large sets of real-time solutions to the Schr\"{o}dinger equation using a realistic in-medium complex-valued potential.  We sample 2 million quarkonia wave packet trajectories and evolve them through the QGP using HQQD to obtain their survival probabilities.  The computation is performed using three different HQQD model parameter sets in order to estimate our systematic uncertainty.  After taking into account final state feed down we compare our results to existing experimental data for the suppression and elliptic flow of bottomonium states and find that HQQD predictions are good agreement with available data for $R_{AA}$ as a function of $N_{\rm part}$ and $p_T$ collected at \mbox{$\sqrt{s_{\rm NN}} =$ 5.02 TeV}.  In the case of $v_2$ for the various states, we find that the path-length dependence of $\Upsilon(1s)$ suppression results in quite small $v_2$ for $\Upsilon(1s)$.  Our prediction for the integrated elliptic flow for $\Upsilon(1s)$ in the $10{-}90$\% centrality class, which now includes an estimate of the systematic error, is \mbox{$v_2[\Upsilon(1s)]$ = 0.003 $\pm$ 0.0007 $\pm\,^{0.0006}_{0.0013}$}.  We also find that, due to their increased suppression, excited bottomonium states have a larger elliptic flow.  Based on this observation we make predictions for $v_2[\Upsilon(2s)]$ and $v_2[\Upsilon(3s)]$ as a function of centrality and transverse momentum.  }


\begin{document}

\setcounter{tocdepth}{1}

\maketitle
\flushbottom

\section{Introduction}
\la{sect:intro}

It is expected that nuclear matter undergoes a phase transition to a primordial state called the quark-gluon plasma (QGP) at temperatures exceeding $T_c \sim 155$ MeV \cite{Bazavov:2013txa,Borsanyi:2016bzg}.  During the phase transition hadronic states breakup and the appropriate degrees of freedom for describing the system become (quasi-) quarks and gluons.  The temperature at which a given hadronic bound state is expected to breakup scales with the binding energy of the state being considered.  Hadronic states with low binding energies such as the $\pi$ are expected to disassociate at temperature close to $T_c$, however, states with higher binding energy such as the $J/\psi$ and $\Upsilon$ have disassociation temperatures well above $T_c$.  Since they are not fully disassociated, one can use the production of heavy quarkonium bound states as a probe of the conditions generated in the QGP, such as its initial central temperature, flow profile, etc.  

Experimentally, the QGP is produced using relativistic heavy-ion collision experiments at Brook-haven National Laboratory's (BNL) Relativistic Heavy Ion Collider (RHIC) and the European Organization for Nuclear Research's (CERN) Large Hadron Collider (LHC).  One key observables for LHC heavy-ion experiments is the production of heavy quarkonium states, with early results measured at the Super Proton Synchrotron (SPS) in the mid-90s already hinting at the QGP-induced suppression of the $J/\psi$.  The expectation that one would observe suppression of heavy-quarkonium states in a thermal medium was based on the early work of Karsch, Mehr, and Satz (KMS) \cite{Matsui:1986dk,Karsch:1987pv,PhysRevD.70.054507,PhysRevC.70.021901} who pointed out using a non-relativistic potential model that in-medium screening of the gluonic interactions would cause heavy-quark bound states to disassociate.  Such a non-relativistic treatment is justified by the fact that, as the mass of the heavy-quark increases, its velocity inside the bound state decreases \cite{Brambilla:2010cs,Brambilla:2010vq,Andronic:2015wma,Mocsy:2013syh}.  This understanding can be made more formal using effective field theory methods to integrate out different energy/momentum scales, resulting in potential non-relativistic QCD (pNRQCD) \cite{Pineda:1997bj,Brambilla:1999xf,Brambilla:2010xn}.    

One key difference from the early days of KMS is that it is now understood that the in-medium heavy-quark potential is complex-valued, with the imaginary part being related to the in-medium breakup rate of heavy-quark bound states \cite{Laine:2006ns,Dumitru:2007hy,Brambilla:2008cx,Burnier:2009yu,Dumitru:2009fy,Dumitru:2009ni,Margotta:2011ta,Du:2016wdx,Nopoush:2017zbu,Guo:2018vwy}.   The imaginary part of the potential was shown to be related to gluon dissociation or parton free dissociation of the states in Refs.~\cite{Brambilla:2011sg,Brambilla:2013dpa}.  As a consequence of the imaginary part of the potential one has a non-Hermitian Hamiltonian and the time evolution of the system becomes non-unitary.  This can be understood in the context of open quantum systems in which the heavy-quark bound state is coupled to a thermal heat bath \cite{Akamatsu:2011se,Akamatsu:2012vt,Akamatsu:2014qsa,Katz:2015qja,Brambilla:2016wgg,Kajimoto:2017rel,Brambilla:2017zei,Blaizot:2017ypk,Blaizot:2018oev,Yao:2018nmy}.  Such imaginary contributions also appear in the context of kinetic transport models since in these calculations one also includes the possibility of in-medium breakup \cite{Grandchamp:2005yw,Rapp:2008tf,Emerick:2011xu,Du:2017qkv,Yao:2018sgn,Du:2019tjf,Yao:2020xzw,Yao:2020xwx}.

In this paper, we follow up our previous work from Ref.~\cite{Islam:2020gdv} by including more details of the calculation, in particular with respect to the numerical method for solving the time-dependent Schr\"odinger equation and our method for including the effect of late-time feed down and we update our treatment of feed down to include a 9 $\times$ 9 feed down matrix.  In addition, in this paper we provide estimates of uncertainties associated with the choice of model parameters used in Ref.~\cite{Islam:2020gdv}.  As in Ref.~\cite{Islam:2020gdv} we will focus on bottomonium states and present a model called Heavy Quarkonium Quantum Dynamics (HQQD) in which we solve the time-dependent Schr\"{o}dinger equation with a complex in-medium potential for a large set of Monte-Carlo-sampled bottomonium wave-packet trajectories.  For each trajectory, the quantum states are in a quantum linear superposition and one can extract the survival probability of a given state by computing the quantum-mechanical overlap of the state's vacuum eigenstate with the in-medium evolved quantum wave-function.  Since we do not include an explicit time-dependent noise contribution in the potential, the resulting wave-functions correspond to the averaged wave-function (thermal expectation value) \cite{Kajimoto:2017rel}.\footnote{See also Sec.~\ref{sec:methodology} for a succinct proof of this statement.}  This approximation allows us to straightforwardly go beyond previous phenomenological works which made use of the adiabatic approximation instead of real-time solutions \cite{Margotta:2011ta,Strickland:2011mw,Strickland:2011aa,Strickland:2012cq,Krouppa:2015yoa,Krouppa:2016jcl,Krouppa:2017jlg,Jaiswal:2017dxp,Bhaduri:2018iwr,Bhaduri:2020lur}.  In a previous paper \cite{Boyd:2019arx}, we made a preliminary investigation of the effects of relaxing the adiabatic approximation, finding that there were potentially important effects on the survival probability of the states.  

Herein, we turn this approach into a more complete phenomenological framework, which can be used for comparisons with experimental data.  To do this, we make use of the output of a 3+1D anisotropic hydrodynamics code which has been tuned to reproduce a large set of soft hadronic observables in $\sqrt{s_{NN}} = 5.02$ TeV collisions \cite{Florkowski:2010cf,Martinez:2010sc,Bazow:2013ifa,Alqahtani:2017jwl,Alqahtani:2017tnq,Alqahtani:2017mhy,Almaalol:2018gjh,Alqahtani:2020paa}.  After computing each state's survival probability, we then take into account late-time feed down of excited states using vacuum branching ratios available from the Particle Data Group \cite{pdg}.  We find that our HQQD results are in quite reasonable agreement with available data given current uncertainties, however, some quantitative differences remain which motivate going beyond the methods used herein to more fully include the effects of in-medium thermal noise, initial production in octet states, and singlet-octet transitions.

The structure of this paper is as follows.  In Sec.~\ref{sec:methodology} we introduce the theoretical tools necessary to derive the in-medium time-dependent Schr\"odinger equation obeyed by heavy-quark bound states.  In Sec.~\ref{sec:potential} we introduce the phenomenological potential model used in HQQD.  In Sec.~\ref{sec:numerics} we provide details of the split-step pseudo-spectral method used herein.  In Sec.~\ref{sec:feeddown} we provide details concerning the implementation of late-time feed-down and the computation of $R_{AA}[\Upsilon]$ and $v_2[\Upsilon]$.  In Sec.~\ref{sec:results} we present our final results, providing estimates of both the statistical errors coming from sampling and systematic errors associated with varying the model parameters.  In Sec.~\ref{sec:conclusions} we give our conclusions and an outlook for the future.

\section{Relation of noise and the imaginary part of the potential}
\label{sec:methodology}

To begin we review the stochastic potential model with two noise sources for a heavy quark and anti-quark described in Ref.~\cite{Akamatsu:2014qsa,Kajimoto:2017rel}.  Let us consider a heavy quark located at $\mathbf{x}=\mathbf{R}+\frac{\mathbf{r}}{2}$ and a heavy anti-quark at $\mathbf{x^\prime}=\mathbf{R}-\frac{\mathbf{r}}{2}$, and the strength of the potential between them is not only weakened by screening but also fluctuates. Therefore, the Hamiltonian consists of a screened potential $V(\mathbf{r})$ and stochastic noise term $\Theta(\mathbf{r},t)$
\begin{equation}
H(\mathbf{r},t)=-\frac{\mathbf{\nabla^2_\mathbf{r}}}{M} + V(\mathbf{r}) +\Theta(\mathbf{r},t) \,\,,
\label{eq:s2.1}
\end{equation}
where,
\begin{equation}
\Theta(\mathbf{r},t)=\theta\left(\mathbf{R}+\frac{\mathbf{r}}{2},t\right) - \theta\left(\mathbf{R}-\frac{\mathbf{r}}{2},t\right)\,\,.
\label{eq:s2.2}
\end{equation}
Here, $\Theta(\mathbf{r},t)$ is the sum of noise terms for the heavy quark, $\theta(\mathbf{x},t)$, and anti-quark, $\theta(\mathbf{x^\prime},t)$, which have the following correlations
\begin{equation}
\langle\theta(\mathbf{x},t)\rangle=0\,\,,\quad \langle\theta(\mathbf{x},t)\theta(\mathbf{x^\prime},t^\prime)\rangle=D(\mathbf{x}-\mathbf{x^\prime})\delta(t-t^\prime)\,\,.
\label{eq:s2.3}
\end{equation}
From Eq.(\ref{eq:s2.2}), we write 
\begin{eqnarray}
\Theta(\mathbf{r},t)\Theta(\mathbf{r^\prime},t^\prime)&=&\bigg[\theta\left(\mathbf{R}+\frac{\mathbf{r}}{2},t\right) - \theta\left(\mathbf{R}-\frac{\mathbf{r}}{2},t\right)\bigg]\bigg[\theta\left(\mathbf{R}+\frac{\mathbf{r^\prime}}{2},t^\prime\right) - \theta\left(\mathbf{R}-\frac{\mathbf{r^\prime}}{2},t^\prime\right)\bigg]\nonumber\\
&=& \theta\left(\mathbf{R}+\frac{\mathbf{r}}{2},t\right)\theta\left(\mathbf{R}+\frac{\mathbf{r^\prime}}{2},t^\prime\right)-\theta\left(\mathbf{R}+\frac{\mathbf{r}}{2},t\right)\theta\left(\mathbf{R}-\frac{\mathbf{r^\prime}}{2},t^\prime\right)\nonumber\\
&& -\theta\left(\mathbf{R}-\frac{\mathbf{r}}{2},t\right)\theta\left(\mathbf{R}+\frac{\mathbf{r^\prime}}{2},t^\prime\right)+\theta\left(\mathbf{R}-\frac{\mathbf{r}}{2},t\right)\theta\left(\mathbf{R}-\frac{\mathbf{r^\prime}}{2},t^\prime\right)\,\,.
\label{eq:s2.4}
\end{eqnarray}
Using Eq.(\ref{eq:s2.3}) in Eq.(\ref{eq:s2.4}) and assuming that the noise correlation $D$-function is an even function i.e., $D(-\mathbf{r})=D(\mathbf{r})$, one obtains the correlation of $\Theta(\mathbf{r},t)$
\begin{eqnarray}
\big\langle\Theta(\mathbf{r},t)\Theta(\mathbf{r^\prime},t^\prime)\big\rangle&=&D\left(\frac{\mathbf{r}-\mathbf{r^\prime}}{2}\right)\delta(t-t^\prime)-D\left(\frac{\mathbf{r}+\mathbf{r^\prime}}{2}\right)\delta(t-t^\prime)\nonumber\\
&&-D\left(-\frac{\mathbf{r}+\mathbf{r^\prime}}{2}\right)\delta(t-t^\prime) +D\left(-\frac{\mathbf{r}-\mathbf{r^\prime}}{2}\right)\delta(t-t^\prime)\nonumber\\
&=&2\bigg[D\left(\frac{\mathbf{r}-\mathbf{r^\prime}}{2}\right)-D\left(\frac{\mathbf{r}+\mathbf{r^\prime}}{2}\right)\bigg]\delta(t-t^\prime)\,\,.
\label{eq:s2.5}
\end{eqnarray}


Next we expand the unitary time evolution operator $e^{-i\Delta t\,H(\mathbf{r},t)}$  in terms of the infinitesimal time $\Delta t$:
\begin{eqnarray}
e^{-i\Delta t\,H(\mathbf{r},t)}&=&1-i\Delta t\,H(\mathbf{r},t)-\frac{1}{2}\big\{\Delta t\,H(\mathbf{r},t)\big \}^2+...\nonumber\\
&=&1-i\Delta t\,H(\mathbf{r},t)-\frac{1}{2}\big\{\Delta t\,\Theta(\mathbf{r},t)\big \}^2+\mathcal{O}\left((\Delta t)^{3/2}\right)\\\label{eq:s2.6}
&\approx&1-i\Delta t\bigg[H(\mathbf{r},t)-\frac{i}{2}\Delta t\Theta(\mathbf{r},t)^2\bigg]\nonumber\\
&\approx&1-i\Delta t\bigg[H(\mathbf{r},t)-\frac{i}{2}\Delta t\bigg\{\theta(\mathbf{x},t)^2+\theta(\mathbf{x^\prime},t)^2-2\,\theta(\mathbf{x},t)\theta(\mathbf{x^\prime},t)\bigg \}\bigg]\,\,,
\label{eq:s2.7}
\end{eqnarray}
where in going from the second to third lines we have dropped the terms of $\mathcal{O}\left((\Delta t)^{3/2}\right)$ and higher.
Here in Eq.~(\ref{eq:s2.7}) we use $\Theta(\mathbf{r},t)=\theta\left(\mathbf{x},t\right) - \theta\left(\mathbf{x^\prime},t\right)\,\,.$ From Eq.~(\ref{eq:s2.7}) we extract the effective Hamiltonian as 
\begin{eqnarray}
\mathrm{H_{eff}}(\mathbf{r},t)&\simeq& H(\mathbf{r},t)-\frac{i}{2}\Delta t\bigg\{\langle\theta(\mathbf{x},t)^2\rangle+\langle\theta(\mathbf{x^\prime},t)^2\rangle-2\,\langle\theta(\mathbf{x},t)\theta(\mathbf{x^\prime},t)\rangle\bigg \} \nonumber\\
&=& H(\mathbf{r},t)-\frac{i}{2}\Delta t\bigg\{D(\mathbf{0})+D(\mathbf{0})-2\,D(\mathbf{x}-\mathbf{x^\prime})\bigg \}\frac{1}{\Delta t}\label{eq:s2.8}\\
&=& H(\mathbf{r},t)-i\bigg\{D(\mathbf{0})-D(\mathbf{x}-\mathbf{x^\prime})\bigg \}\nonumber\\
&=& H(\mathbf{r},t)-i\bigg\{D(\mathbf{0})-D(\mathbf{r})\bigg \}\label{eq:s2.9}\\
&=&-\frac{\nabla^2_\mathbf{r}}{M} + V(\mathbf{r})-i\bigg\{D(\mathbf{0})-D(\mathbf{r})\bigg \} +\Theta(\mathbf{r},t)\,\, ,
\label{eq:s2.10}
\end{eqnarray}
where in Eq.~(\ref{eq:s2.8}), we use $\langle\theta(\mathbf{x},t)\theta(\mathbf{x^\prime},t^\prime)\rangle=D(\mathbf{x}-\mathbf{x^\prime})\frac{\delta_{tt^\prime}}{\Delta t}\,\,,$ in Eq.~(\ref{eq:s2.9}) we use the fact that $\mathbf{x}=\mathbf{R}+\frac{\mathbf{r}}{2}$, $\mathbf{x^\prime}=\mathbf{R}-\frac{\mathbf{r}}{2}\Rightarrow \mathbf{x}-\mathbf{x^\prime}=\mathbf{r}\,\,,$ and, in Eq.~(\ref{eq:s2.10}), we have inserted Eq.~(\ref{eq:s2.1}).

Since $\langle \Theta(\mathbf{r},t)\rangle=0$, the averaged effective Hamiltonian can be written as 
\begin{equation}
\langle\mathrm{H_{eff}}(\mathbf{r},t)\rangle=-\frac{\nabla^2_\mathbf{r}}{M} + V(\mathbf{r})-i\bigg\{D(\mathbf{0})-D(\mathbf{r})\bigg \}\,\,.
\label{eq:s2.11}
\end{equation}
Therefore, the stochastic Schr\"odinger equation for the noise averaged quarkonium wave function is,
\begin{eqnarray}
i\frac{\partial}{\partial t}\langle\Psi_{Q\bar{Q}}(\mathbf{r},t)\rangle&=&\langle\mathrm{H_{eff}}(\mathbf{r},t)\rangle\langle\Psi_{Q\bar{Q}}(\mathbf{r},t)\rangle\nonumber\\
&=&\bigg[-\frac{\nabla^2_\mathbf{r}}{M} + V(\mathbf{r})-i\Big\{D(\mathbf{0})-D(\mathbf{r})\Big \}\bigg]\langle\Psi_{Q\bar{Q}}(\mathbf{r},t)\rangle\,\,.
\label{eq:s2.12}
\end{eqnarray}
From this expression, it can be seen that the imaginary part of the potential is related to the $D$-function via $\Im[V(r)]=D(\mathbf{r})-D(\mathbf{0})$.

\section{Phenomenological potential}
\label{sec:potential}

In vacuum we take the heavy-quark potential to be given by a Cornell potential with a finite string breaking distance
\be
V_{\rm vac}(r) =
\begin{cases}  
	-\frac{a}{r} + \sigma r &\mbox{if } r \leq r_{\rm SB} \\
	-\frac{a}{r_{\rm SB} } + \sigma r_{\rm SB}   & \mbox{if } r > r_{\rm SB}
\end{cases} \, ,
\label{eq:vvac}
\ee
where \mbox{$a = 0.409$} is the effective coupling, \mbox{$\sigma = 0.21$~GeV$^2$} is the string tension, and \mbox{$r_{\rm SB}  =$ 1.25 fm} is the string breaking distance.  Using this set of vacuum parameters and assuming \mbox{$M_b = 4.7$ GeV} one obtains vacuum masses of \mbox{$\{9.46,10.0,9.88,10.36,10.25,10.13\}$ GeV} for $\Upsilon(1s)$, $\Upsilon(2s)$, $\chi_b(1p)$, $\Upsilon(3s)$, and $\chi_b(2p)$, respectively.

For the in-medium potential we use an internal energy based potential.   The free energy of a static heavy quark-antiquark pair in an isotropic plasma \cite{Matsui:1986dk,Karsch:1987pv,PhysRevD.70.054507,PhysRevC.70.021901,Strickland:2011aa}
\begin{eqnarray}
F(r,t) &=& -\frac{g^2 C_F}{4\pi r}e^{-m_D r}+\frac{\sigma}{m_D}\left[1-e^{-m_D r}\right]\nonumber\\
&=& -\frac{a}{r}e^{-m_D r}+\frac{\sigma}{m_D}\left[1-e^{-m_D r}\right]\quad ,
\label{eq:2.4}
\end{eqnarray}
where $a=g^2 C_F/4\pi$ and $\sigma$ is the string tension. We take the in-medium gluonic Debye mass to be $m_D = \lambda \sqrt{4 \pi N_c (1 + N_f/6) \alpha_s T^2/3}$.  Here we will consider the cases $\lambda \in \{0.8,1,1.2\}$, with the central value being the known high-temperature limit of the gluonic Debye mass.

The internal energy, $U$, of the states which has an entropy contribution is given by
 \begin{eqnarray}
 U&=& F+TS \nonumber\\
 &=&-\frac{a}{r}(1+m_D r)e^{-m_D r}+\frac{2\sigma}{m_D}\left[1-e^{-m_D r}\right]-\sigma r e^{-m_D r} \, ,
 \label{eq:2.7}
 \end{eqnarray}
where we have used $S=-\frac{\partial F}{\partial T}$.
Eq.\eqref{eq:2.7} gives us the internal energy based real part of the Karsch-Mehr-Satz (KMS) potential

\begin{equation}
V_{\mathrm{KMS}}(r)=U=-\frac{a}{r}(1+m_D r)e^{-m_D r}+\frac{2\sigma}{m_D}\left[1-e^{-m_D r}\right]-\sigma r e^{-m_D r} \, .
\label{eq:2.8}
\end{equation}

We note that, in equilibrium, there are arguments to support the use of the internal energy~\cite{PhysRevD.70.054507,PhysRevC.70.021901,SHURYAK200564}, however, since bottomonium states are not expected to be in thermal equilibrium with their surroundings, it is unclear what the correct choice should be.  In order to make sure that we have the correct equilibrium limit we choose to use the internal energy prescription.  To match smoothly onto the zero temperature limit we use
\be
V_R(r)=\Re[V(r)] =
\begin{cases}  V_{\rm KMS}(r)  &\mbox{if } V_{\rm KMS}(r)  \leq V_{\rm vac}(r_{\rm SB}) \\
	V_{\rm vac}(r_{\rm SB}) & \mbox{if } V_{\rm KMS}(r) > V_{\rm vac}(r_{\rm SB})
\end{cases} \, .
\label{eq:vmedre}
\ee
In the limit $T\rightarrow0$, Eq.~\eqref{eq:vmedre} reduces to Eq.~\eqref{eq:vvac}.

For the imaginary part of the potential we take the result of the leading-order resummed perturbative QCD calculation of Laine et al 
\be
V_I(r)=\Im[V(r)] = - C_F \alpha_s T \phi(m_D r) \, ,
\label{eq:vmedim}
\ee
with $\phi(\hat{r}) \equiv 1 - 2 \int_0^\infty \sin(z)/(z^2 + \hat{r}^2)^2 $ \cite{Laine:2006ns}.  We evaluate the strong coupling $\alpha_s$ at the scale $\mu = 2 \pi T$ and use three-loop running \cite{pdg} with \mbox{$\Lambda_{\overline{MS}} = 344$ MeV}.  This value of $\Lambda_{\overline{MS}}$ is chosen in order to reproduce the lattice result for the running coupling $\alpha_s(5\text{ GeV}) = 0.2034$~\cite{McNeile:2010ji}.
The resulting final complex-valued potential is of the form 
\begin{equation}
V(r) = V_R(r) + i V_I(r) \, .
\label{eq:vform}
\end{equation}

\section{Numerical method for solving the Schr\"odinger equation}
\label{sec:numerics}

The complex potential is a purely radial potential hence the general solution in spherical coordinates can be written as
\be
\psi(r,\theta,\phi,t) = \sum_{\ell m} R_{\ell m}(r,t) Y_{\ell m}(\theta,\phi) \, ,
\ee
where $Y_{\ell m}$ are spherical harmonics. Making a change of variables to \mbox{$u_{\ell m}(r,t) \equiv r R_{\ell m}(r,t)$}, we obtain 
\be
u(r,\theta,\phi,t) = \sum_{\ell m} u_{\ell m}(r,t) Y_{\ell m}(\theta,\phi) \, ,
\ee
where $u(r,\theta,\phi,t) = r \psi(r,\theta,\phi,t)$. This change of variables allows us to write the Hamiltonian in one-dimensional form
\be
\hat{H}_\ell = \frac{\hat{p}^2}{2m} + V_{{\rm eff},\ell}(r,t) \, ,
\ee
where $V_{{\rm eff},\ell}(r,t) = V(r,t) + \frac{\ell(\ell+1)}{2 m r^2}$ and $\hat{p} = -i \frac{d}{dr}$.

\paragraph*{}

By applying the time evolution operator to $u$, one obtains \cite{Boyd:2019arx}
\ba
u(r,\theta,\phi,t + \Delta t) &=& \exp(- i \hat{H} \Delta t) u(r,\theta,\phi,t ) \nonumber \\
&=&  {\cal N} \sum_{\ell,m} \frac{1}{\sqrt{2\ell+1}} Y_{\ell m}(\theta,\phi) \exp(- i \hat{H}_\ell \Delta t) u_\ell(r,t)  \, ,
\ea
where ${\cal N}$ is a normalization constant, and the summation limits are implicit. 
Using $u_{\ell}(r,t) = \sqrt{2\ell+1} \int d\Omega \, Y_{\ell m}^*(\theta,\phi) \, u(r,\theta,\phi,t)$, we obtain
\be
u_\ell(r,t + \Delta t) = \exp(- i \hat{H}_\ell \Delta t) u_\ell(r,t) \, ,
\label{eq:uUpdate}
\ee
which tells us that each of the different $\ell$ states can be updated independently using the corresponding Hamiltonian.  Finally, for the normalization of the various states, we take
\be
\int_0^\infty dr \, u_\ell^*(r,t) u_\ell(r,t) = 1 \, .
\ee
\paragraph*{}

We will use the time evolution operator \eqref{eq:uUpdate} to evolve the wave function given a particular initial condition.  To enforce the fact that the function $u_\ell(r,t)$ must vanish at the origin, we use real-valued Fourier sine series to describe both the real and imaginary parts of the wave function.  Hence, the correct boundary conditions at $r=0$ are satisfied automatically.  The resulting time evolution steps are \cite{Boyd:2019arx}


\begin{itemize}
	\item Step-1: Update in configuration space using a half-step: $ \psi_1 = \exp(- i V \Delta t/2) \psi_0$.
	
	\item Step-2: Perform Fourier sine transformations ($\mathbb{F}_s$) on real and imaginary parts separately: $\tilde\psi_1 = \mathbb{F}_s[\Re \psi_1] + i \mathbb{F}_s[\Im \psi_1] $.
	
	\item Step-3: Update in momentum space using: $\tilde\psi_2 =  \exp\!\left(-i \frac{p^2}{2 m} \Delta t\right) \tilde\psi_1$.
	
	\item Step-4: Perform inverse Fourier sine transformations ($\mathbb{F}_s^{-1}$) on real and imaginary parts separately: $\psi_2 = \mathbb{F}_s^{-1}[\Re \tilde\psi_2] + i \mathbb{F}_s^{-1}[\Im \tilde\psi_2] $.
	
	\item Step-5: Update in configuration space using a half-step: $ \psi_3 = \exp(- i V \Delta t/2) \psi_2$.
\end{itemize}

The discrete sine transforms (DST) above can be implemented using standard routines for Fast Fourier Transforms.  By repeating this procedure, we can evolve the wave function forward in time in a manifestly unitary manner which is generally faster and more accurate than traditional finite-difference based evolution schemes.  We note that besides the improved performance observed, one major benefit of the DST algorithm is that the derivatives are computed using all points in the lattice, not just at just fixed number of points. As a result, the evolution obtained using the DST algorithm is more accurate than with, for example, a Crank-Nicolson (CN) scheme using a three-point derivative \cite{Boyd:2019arx}.  In practice, we use the CUDA CUFFT library for the implementation \cite{cuda}.  The resulting code is massively parallelized.

\section{Computation of $R_{AA}$ including feed-down}
\label{sec:feeddown}

Using the complex potential Eq.~\eqref{eq:vform} along with Eq.~\eqref{eq:vmedre} and Eq.~\eqref{eq:vmedim}, we then numerically solve the time-dependent Schr\"odinger equation on a discrete lattice. The DST method used is manifestly unitary for real-valued potentials and based on a split-step pseudospectral method described above \cite{Fornberg:1978,TAHA1984203,Boyd:2019arx}. For the discretization we use $N = 4096$ points in $r$ with $L = r_{\rm max} = 19.7$ fm with a corresponding lattice spacing of $a \simeq 0.0048$ fm.  We compute the in-medium suppression for $\ell=0$ and $\ell=1$ states, separately, following Eq.~\eqref{eq:uUpdate}.

We choose a Gaussian initial wave-function \cite{Islam:2020gdv}
\be
u_\ell(r,\tau=0) \propto r^{\ell+1} \exp(-r^2/\Delta^2) \, ,
\ee
with $\Delta = 0.04$ fm. This initial conditions mimics the local (delta function) production of bottomonium states in space resulting from their initial hard production.  This type of initial state is a quantum superposition of many eigenstates of the Schr\"odinger equation for a fixed $\ell$. Once the wave-function evolves forward in time, we obtain the survival probability of a given vacuum state by computing the overlap of the in-medium quantum wave-function with the vacuum basis states. These overlaps decay in time because of the non-Hermitian nature of the system's Hamiltonian which is fundamentally related to the in-medium breakup of bottomonium states. 

Here we solve the 3+1D Schr\"odinger equation for a large set of trajectories using a realistic 3+1D hydrodynamics background tuned to data~\cite{Alqahtani:2020paa,Alqahtani:2017mhy}.  We use smooth optical Glauber initial conditions which provide the initial energy density profile as a function of the impact parameter.  For the background evolution we used an initial central temperature of \mbox{$T_0 = 630$ MeV} at \mbox{$\tau_0 = 0.25$ fm/c} and a constant specific shear viscosity of \mbox{$4\pi\eta/s = 2$}, which correspond to the results obtained in a recent aHydroQP fit to soft observables at 5 TeV~\cite{Alqahtani:2020paa}. We evolve the quantum wave-packets using the vacuum potential starting at $\tau = 0$ fm/c and turn on the in-medium complex potential at $\tau = \tau_\text{med}$. Finally, when the temperature on a given trajectory drops below the QGP transition temperature, \mbox{$T_\text{QGP} = 155$ MeV}, we again use the vacuum potential for the wave-function evolution.  In the results section we will present our findings when varying $\tau_\text{med} \in \{ 0.25, 0.4, 0.6 \}$ fm/c

Due to the fact that each sampled wave-packet experiences a different temperature along its trajectory while propagating through the QGP, we numerically solve the time-dependent Schr\"odinger equation for a large set of bottomonium trajectories (2 million). We assume that the initial transverse spatial distribution for bottomonium production is proportional to the binary overlap profile of the two colliding nuclei, $N_{AA}^\text{bin}(x,y)$ and then use Monte-Carlo sampling to generate the initial production points. Here, we consider all bottomonia to have zero rapidity, $y=0$, in order to simplify the calculation in a manner that takes advantage of the approximate boost-invariance of the QGP. We assume that the transverse momentum ($p_T$)-distribution is proportional to $p_T/(p_T^2 + \langle M \rangle^2)^2$ for all the states, where $\langle M \rangle$ is the average mass of all states being considered and, once again, we Monte-Carlo sample the $p_T$ for each particle generated.  Finally we sample the initial azimuthal angle $\phi$ from a uniform distribution between 0 and $2\pi$. Once we have the initial position, transverse momentum, and azimuthal angle, we then record the aHydroQP result for the QGP temperature along the (assumed) straight line trajectory followed by the quantum wave-packets.

\begin{table}[t]
	\centering
	\includegraphics[scale=.77]{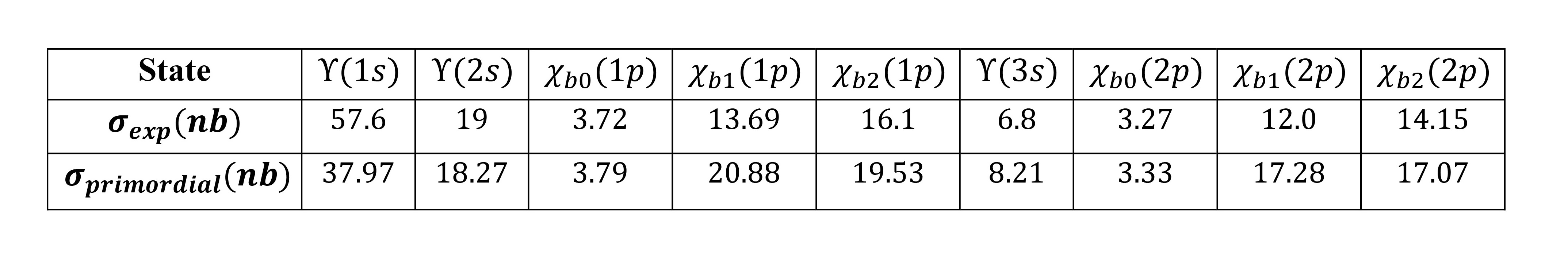}
	\caption{The experimentally observed ($\sigma_{exp}$) and primordial ($\sigma_{primordial}$) $pp \rightarrow$ bottomonium production cross sections of the various bottomonium states.  Data were collected from Refs.~\cite{pdg,Khachatryan:2016xxp,CMS5TeV,Sirunyan:2018nsz,Aaij:2014caa,HeeSok,Khachatryan:2014ofa,Lansberg:2019adr}. The primordial (direct) cross-sections were obtained from the experimentally measured post feed down cross sections using, $\vec{\sigma}_\text{primordial} = F^{-1} \vec{\sigma}_\text{exp}$.}
	\label{tab:01new}
\end{table}

Once each quantum state propagates along its trajectory, the survival probabilities are converted into particle number by multiplying by 
\begin{enumerate}
	\item The expected number of binary collisions in the centrality bin sampled and
	\item The primordial (direct) production cross section for each bottomonium state.
\end{enumerate}

At this point one needs to take into account the late time feed down of excited bottomonium states.  This can be done by introducing a feed down matrix $F$ which collects the known information about excited bottomonium state decays available from the Particle Data Group \cite{pdg}.  In the case of proton-proton collisions one can take the primordial cross sections for production and convert this into the post feed down cross sections (abundances) by applying a feed down matrix to the primordial cross sections $\vec{\sigma}_\text{exp} = F \vec{\sigma}_\text{primordial}$ 
\be
	F = \left(
	\begin{array}{ccccccccc}
		1 & 0.2645 & 0.0194 & 0.352 & 0.18 & 0.0657 & 0.0038 & 0.1153 & 0.077 \\
		0 & 0.7355 & 0 & 0 & 0 & 0.106 & 0.0138 & 0.181 & 0.089 \\
		0 & 0 & 0.9806 & 0 & 0 & 0 & 0 & 0 & 0 \\
		0 & 0 & 0 & 0.648 & 0 & 0 & 0 & 0.0091 & 0 \\
		0 & 0 & 0 & 0 & 0.82 & 0 & 0 & 0 & 0.0051 \\
		0 & 0 & 0 & 0 & 0 & 0.8283 & 0 & 0 & 0 \\
		0 & 0 & 0 & 0 & 0 & 0 & 0.9824 & 0 & 0 \\
		0 & 0 & 0 & 0 & 0 & 0 & 0 & 0.6946 & 0 \\
		0 & 0 & 0 & 0 & 0 & 0 & 0 & 0 & 0.8289 \\
	\end{array}
	\right) ,
	\label{eq:fdmnew}
	\ee
with the vector $\vec{\sigma}$ collecting the observed and primordial (direct) cross sections for the $\{ \Upsilon(1s), \Upsilon(2s), \chi_{b0}(1p), \chi_{b1}(1p),\chi_{b2}(1p),\Upsilon(3s), \chi_{b0}(2p),\chi_{b1}(2p),\chi_{b2}(2p) \}$. Knowing the experimental values for the production cross-sections $\vec{\sigma}_\text{exp}$ (listed in Tab.~\ref{tab:01new}), one can compute the primordial cross sections via $\vec{\sigma}_\text{primordial} = F^{-1} \vec{\sigma}_\text{exp}$.  The resulting values for $\vec{\sigma}_\text{primordial}$ are listed in Tab.~\ref{tab:01new}.

We extract the cross sections of $\Upsilon(1s)$,  $\Upsilon(2s)$, and $\Upsilon(3s)$ from the 5.02 TeV data obtained by the CMS collaboration in the rapidity interval $|y| \leq 2.4$ \cite{Sirunyan:2018nsz}. The left panel of figure 3 of Ref.~\cite{Sirunyan:2018nsz} presents $B \times d\sigma/dy$, where $B$ is the dimuon branching fraction. Once we take the average over rapidity in the interval presented in that CMS figure, we obtain $B \times d\sigma/dy\approx$ 1.44 nb, 0.37 nb, and 0.15 nb, respectively. Dividing by the branching fractions for $\Upsilon(1s)$,  $\Upsilon(2s)$, and $\Upsilon(3s)$ $\rightarrow \mu^+ \mu^-$, which are $\approx$ 2.5\%, 1.9\%, and 2.2\% \cite{pdg}, respectively \cite{pdg}, and obtains
	\begin{equation}
	\langle d\sigma[\Upsilon(1s), \Upsilon(2s), \Upsilon(3s)]/dy \rangle_y =  \{57.6, 19, 6.8 \} \text{ nb} \, .
	\end{equation}
The $\chi_{b}$-cross sections can be obtained by through knowledge of $\sigma[\Upsilon(1s)]$, the ratios \\$\sigma[\chi_{bj}(np)]/\sigma[\chi_{bj'}(np)]$, and the ratios of the 1p and 2p cross-sections to 1s cross sections provided in Ref.~\cite{Aaij:2014caa}. We assume  $\sigma[\chi_{b2}(np)]/\sigma[\chi_{b1}(np)] = 1.176$ \cite{HeeSok} for both the $1p$ and $2p$ states.  This value is consistent with available experimental data \cite{Khachatryan:2014ofa}.  To obtain the $\chi_{bj'}(np)$ cross sections, the values of $\mathcal{R}^{\chi_b(nP)}_{\Upsilon(1s)}$ are taken from the lowest $p_{T}$ bins of tables 5 and 6 of Ref.~\cite{Aaij:2014caa} (measured at $\sqrt{s}=7$ TeV and $\sqrt{s}=8$ TeV, respectively) and extrapolated to $\sqrt{s}=5$ TeV.  Using this extrapolation, $\sigma[\chi_{bj}(np)]$ for $j,\, n=1,\,2$ can be extracted by using Eq.~(1) of Ref.~\cite{Aaij:2014caa}. The $\chi_{b0}(np)$ cross sections are taken to be one quarter of the average of the $\sigma[\chi_{b1}(np)]$ and $\sigma[\chi_{b2}(np)]$ cross-sections~\cite{HeeSok}.  
	This gives 
	\begin{align}
	\langle d\sigma[\chi_{b0}(1p),\chi_{b1}(1p),\chi_{b2}(1p)]/dy \rangle_y &= \{ 3.72, 13.69, 16.1 \} \text{ nb} \, , \\
	\langle d\sigma[\chi_{b0}(2p),\chi_{b1}(2p),\chi_{b2}(2p)]/dy \rangle_y &= \{ 3.27, 12.0, 14.15 \} \text{ nb} \, ,
	\end{align}
Note that herein we ignore possible $p_T$-dependence in the branching ratios which could potentially be important for quantitative understanding of the $p_T$-dependence of post feed down $R_{AA}$ \cite{Lansberg:2019adr}.

\begin{table}[t]
	\centering
	\includegraphics[scale=.77]{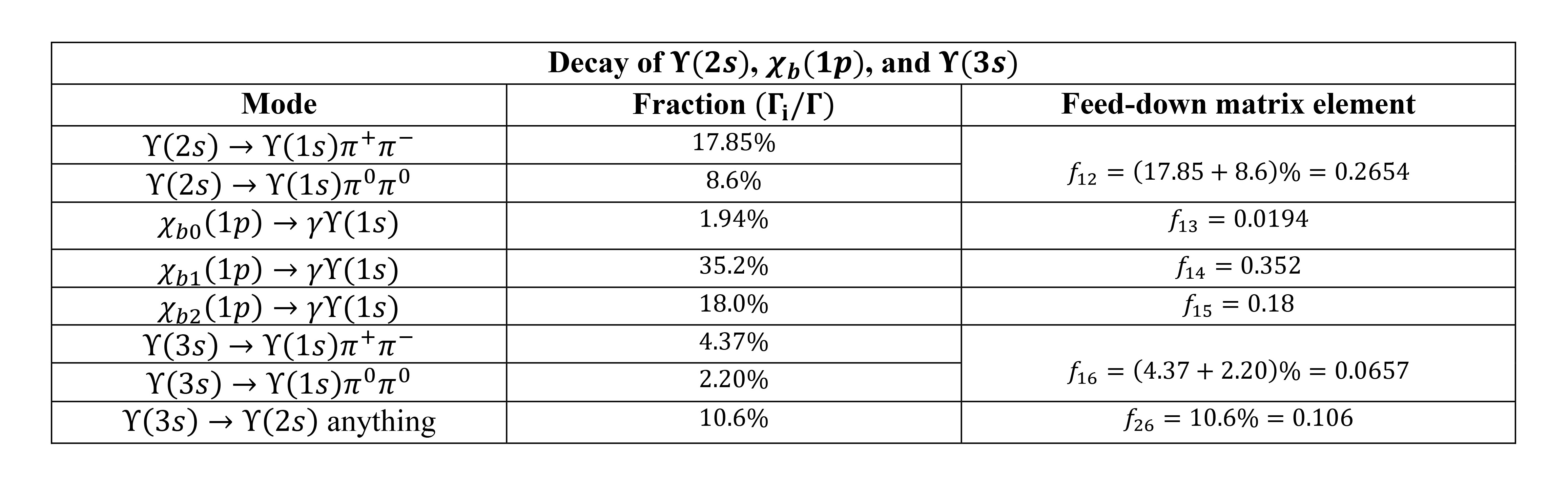}
	\caption{Decay modes of $\Upsilon(2s)$, $\chi_b(1p)$, and $\Upsilon(3s)$ and the calculation of different elements of the feed-down matrix \eqref{eq:fdmnew} using the experimentally measured branching ratios of the various bottomonium states \cite{pdg}.}
	\label{tab:02new}
\end{table}

To compute the effect of final-state feed-down in heavy-ion collisions, we first construct a vector $\vec{N}_\text{QGP}$ containing the numbers of each state produced at the end of each simulation (survival probability $\times \langle N_\text{bin}(b) \rangle \times  \vec{\sigma}_\text{primordial}$).  We then multiply the result by the same feed-down matrix used for pp feed down, i.e. $\vec{N}_\text{final} = F \vec{N}_\text{QGP}$.  The use of the same feed down matrix in both cases is related to the fact that feed down occurs on a time scale of $10^{-20}-10^{-18}$ seconds which is 100-10000 times longer than the QGP lifetime.

The calculation of different elements of the feed-down matrix \eqref{eq:fdmnew} is demonstrated in Table \ref{tab:02new} and Table \ref{tab:03new}. Since each column of the feed-down matrix \eqref{eq:fdmnew}, $F$, must sum to unity in order to preserve bottom number, one can obtain the diagonal elements by using $f_{ii}=1-\sum_j f_{ji}$, where $\{i,j=1,2,3,4,5,6,7,8,9 \}$.\footnote{Here we have neglected decay processes with small branching ratios, such as $\Upsilon(2s) \rightarrow \Upsilon(1s) + \pi^0$, which has a branching ratio of BR $< 4 \times 10^{-5}$ \cite{pdg}.}  Using \eqref{eq:fdmnew}, the final number of different states produced can therefore be computed as follows
\begin{eqnarray}
	N^{\Upsilon(1s)}_\text{final} &=& f_{11} N^{\Upsilon(1s)}_\text{QGP} + f_{12} N^{\Upsilon(2s)}_\text{QGP} +  f_{13} N^{\chi_{b0}(1p)}_\text{QGP} +  f_{14} N^{\chi_{b1}(1p)}_\text{QGP} +  f_{15} N^{\chi_{b2}(1p)}_\text{QGP}\\\nonumber
	&+& f_{16} N^{\Upsilon(3s)}_\text{QGP} +  f_{17} N^{\chi_{b0}(2p)}_\text{QGP} +  f_{18} N^{\chi_{b1}(2p)}_\text{QGP} +  f_{19} N^{\chi_{b2}(2p)}_\text{QGP} \, , \\
	N^{\Upsilon(2s)}_\text{final} &=& f_{22} N^{\Upsilon(2s)}_\text{QGP} + f_{26} N^{\Upsilon(3s)}_\text{QGP} +  f_{27} N^{\chi_{b0}(2p)}_\text{QGP} +  f_{28} N^{\chi_{b1}(2p)}_\text{QGP} +  f_{29} N^{\chi_{b2}(2p)}_\text{QGP} \, ,\\
	N^{\chi_{b0}(1p)}_\text{final} &=& f_{33} N^{\chi_{b0}(1p)}_\text{QGP} \, ,\\
	N^{\chi_{b1}(1p)}_\text{final} &=& f_{44} N^{\chi_{b1}(1p)}_\text{QGP} + f_{48} N^{\chi_{b1}(2p)}_\text{QGP} \, ,\\
	N^{\chi_{b2}(1p)}_\text{final} &=& f_{55} N^{\chi_{b2}(1p)}_\text{QGP} + f_{59} N^{\chi_{b2}(2p)}_\text{QGP} \, ,\\
	N^{\Upsilon(3s)}_\text{final} &=& f_{66} N^{\Upsilon(3s)}_\text{QGP} \, ,\\
	N^{\chi_{b0}(2p)}_\text{final} &=& f_{77} N^{\chi_{b0}(2p)}_\text{QGP} \, ,\\
	N^{\chi_{b1}(2p)}_\text{final} &=& f_{88} N^{\chi_{b1}(2p)}_\text{QGP} \, ,\\
	N^{\chi_{b2}(2p)}_\text{final} &=& f_{99} N^{\chi_{b2}(2p)}_\text{QGP} \, ,
	\end{eqnarray}
where we have discarded terms for which the coefficients were zero in \eqref{eq:fdmnew}.

\begin{table}[t]
	\centering
	\includegraphics[scale=.77]{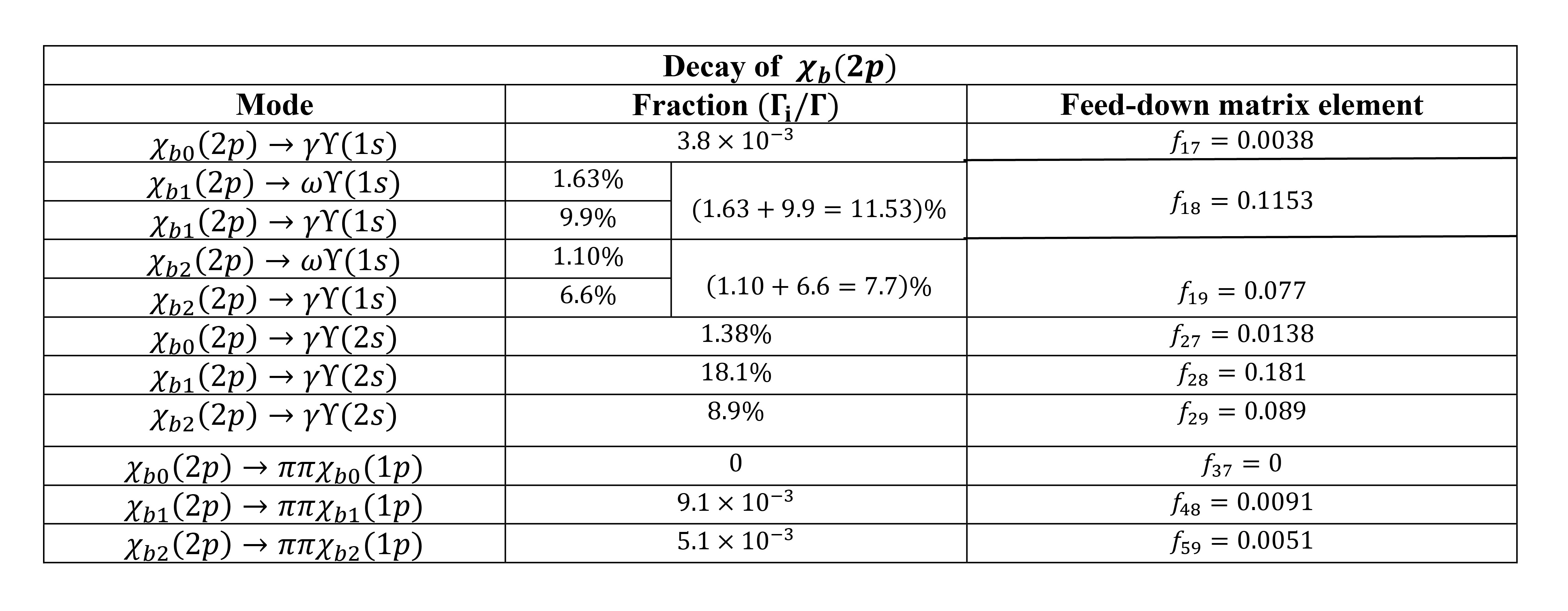}
	\caption{Decay modes of $\chi_b(2p)$ and the calculation of different elements of the feed-down matrix \eqref{eq:fdmnew} using the experimentally measured branching ratios of the various bottomonium states \cite{pdg}.}
	\label{tab:03new}
\end{table}

Finally, we compute $R_{AA}$ for each state by dividing the final number produced after feed down by the average number of binary collisions in the sampled centrality class times the post feed-down $pp$ production cross-section for each state ($\sigma_\text{exp}$).  The momentum anisotropic flow of bottomonium states can be obtained in a similar manner.  In this case, we can exploit the fact that the aHydroQP background always has the reaction plane aligned with the $x$ and $y$ axes, i.e. $\Psi_\text{RP} = 0$, to simply average $\cos(n\phi)$ over all particles produced after feed down is taken into account.  In practice we compute the following average over all sampled trajectories in a given centrality and $p_T$ bin, $v_n \equiv \langle \cos(n\phi) \rangle$.  For both $R_{AA}$ and $v_2$, we can directly compute the statistical uncertainty associated with the sum over the sampled quantum wave-packet trajectories.  To obtain an estimate of our systematic uncertainty we will vary the HQQD model parameters used in Ref.~\cite{Islam:2020gdv}.

\section{Results}
\label{sec:results}

In order to estimate the systematic uncertainty coming from the choice of model parameters we will vary the medium initialization time ($\tau_\text{med}$) and the pre-factor of the peturbative Debye mass ($\lambda$) in the set
\be
(\tau_\text{med},\lambda) \in \{ (0.25 \text{ fm/c},1.2),(0.4 \text{ fm/c},1), (0.6 \text{ fm/c},0.8) \} \, ,
\label{eq:bounds}
\ee
where the factor $\lambda$ accounts for higher-order corrections to the perturbative Debye mass
\be
m_D = \lambda \sqrt{4 \pi N_c (1 + N_f/6) \alpha_s T^2/3} \, ,
\ee
with $\lambda = 1$ giving the known leading-order result from hard-thermal-loop resummation \cite{Andersen:2004fp,Ghiglieri:2020dpq}.

\begin{figure}[t]
	\begin{center}
		\includegraphics[width=0.475\linewidth]{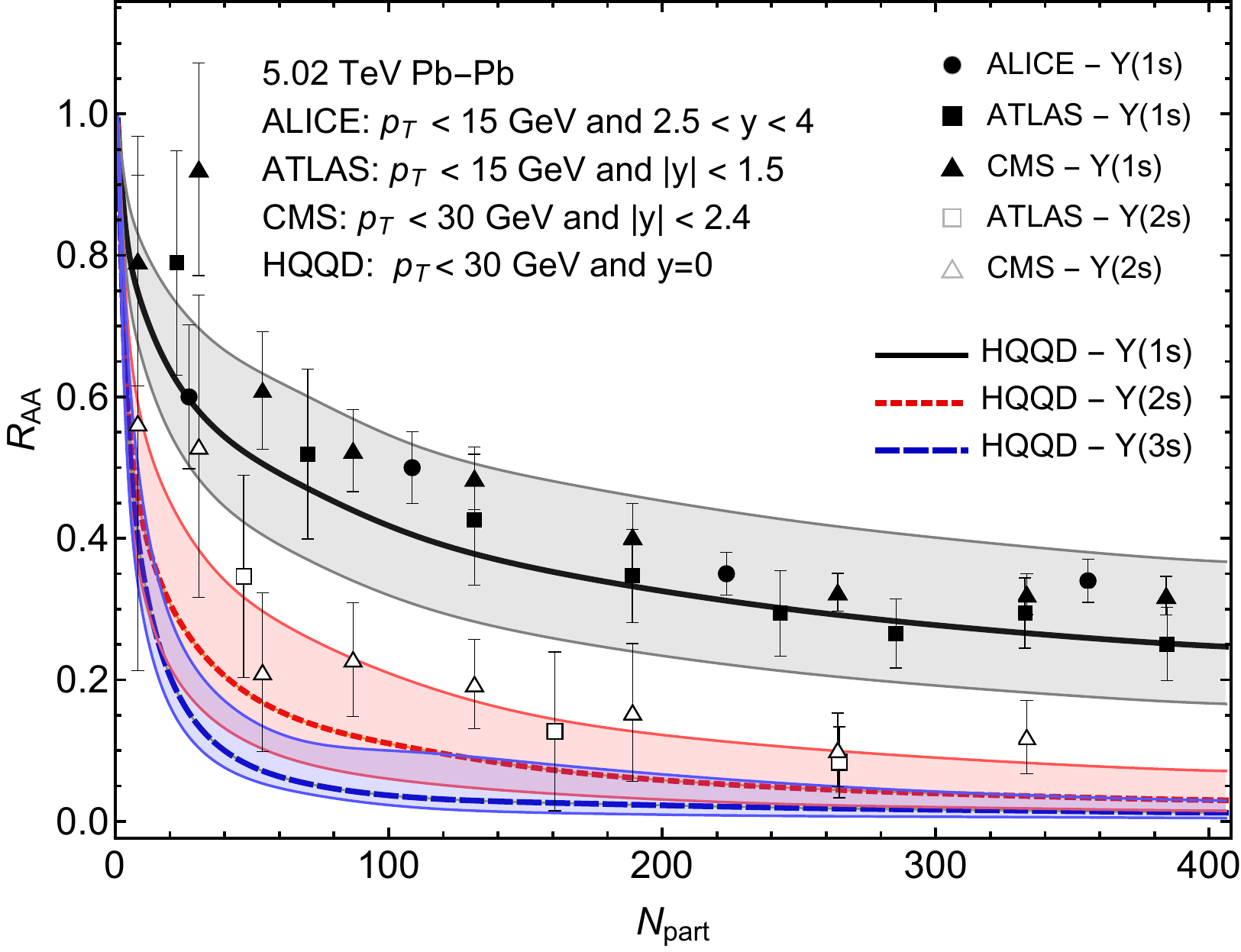} $\;\;\;\;$
		\includegraphics[width=0.475\linewidth]{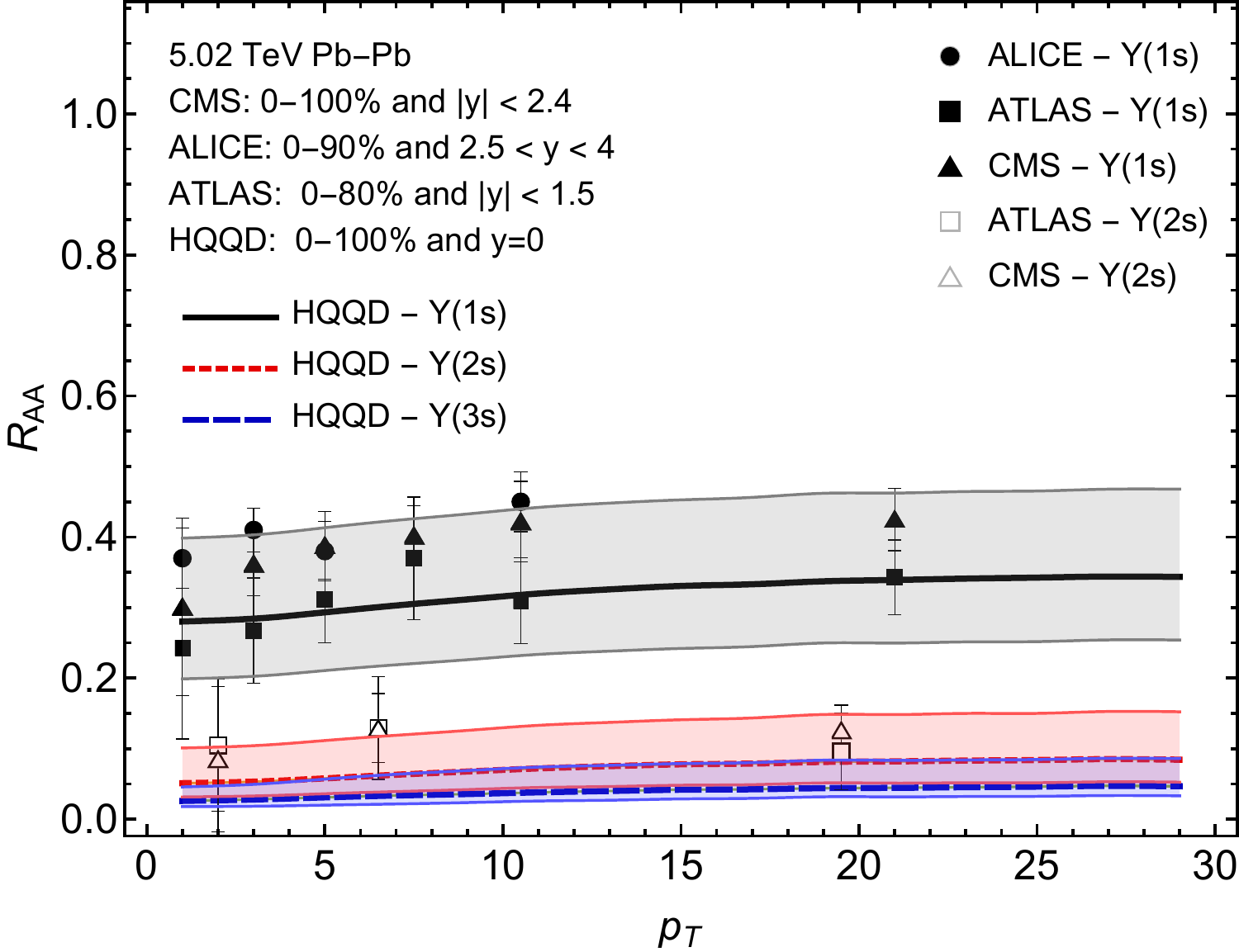}
	\end{center}
	\caption{Nuclear suppression factor, $R_{AA}$, of bottomonium $s$-wave states as a function of $N_\text{part}$ (left) and $p_T$ (right).  The solid, short-dashed, and dashed lines show the predictions of HQQD.  The shaded bands indicate the variation of the HQQD result when changing the model parameters in the range specified in Eq.~\eqref{eq:bounds}.  The data points are from the  ALICE \cite{Acharya:2018mni}, ATLAS \cite{ATLAS5TeV}, and CMS \cite{Sirunyan:2018nsz} collaborations.  Experimental error bars shown were obtained by adding statistical and systematic uncertainties in quadrature.}
	\label{fig:raavsnpartandpt} 
\end{figure}

We present the HQQD predictions for the suppression of $\Upsilon(1s)$, $\Upsilon(2s)$, and $\Upsilon(3s)$ states as a function of $N_{\rm part}$ in Fig.~\ref{fig:raavsnpartandpt} (left).  For the left panel of Fig.~\ref{fig:raavsnpartandpt} we applied a transverse momentum cut of $p_T < 30$ GeV in HQQD.  The HQQD results are compared with experimental data obtained by the ALICE \cite{Acharya:2018mni}, ATLAS \cite{ATLAS5TeV}, and CMS \cite{Sirunyan:2018nsz} collaborations, shown as circles, squares, and triangles, respectively.  From Fig.~\ref{fig:raavsnpartandpt} (left), we see that, for the central choice of model parameters corresponding to $\tau_0 = 0.4\;$fm/c and $\lambda = 1$, HQQD does a quite reasonable job in describing the $N_\text{part}$ dependence of $R_{AA}[\Upsilon(1s)]$.  The shaded bands in Fig.~\ref{fig:raavsnpartandpt} (left) show the variation of the results obtained when taking the limits indicated in Eq.~\eqref{eq:bounds}, with the first and third cases mapping to stronger and weaker suppression compared to the central line, respectively.  In the case of the $\Upsilon(2s)$, one sees that using the central parameters, HQQD seems to under-predict the observed $R_{AA}[\Upsilon(2s)]$.  

\begin{figure}[t]
	\begin{center}
		\includegraphics[width=0.47\linewidth]{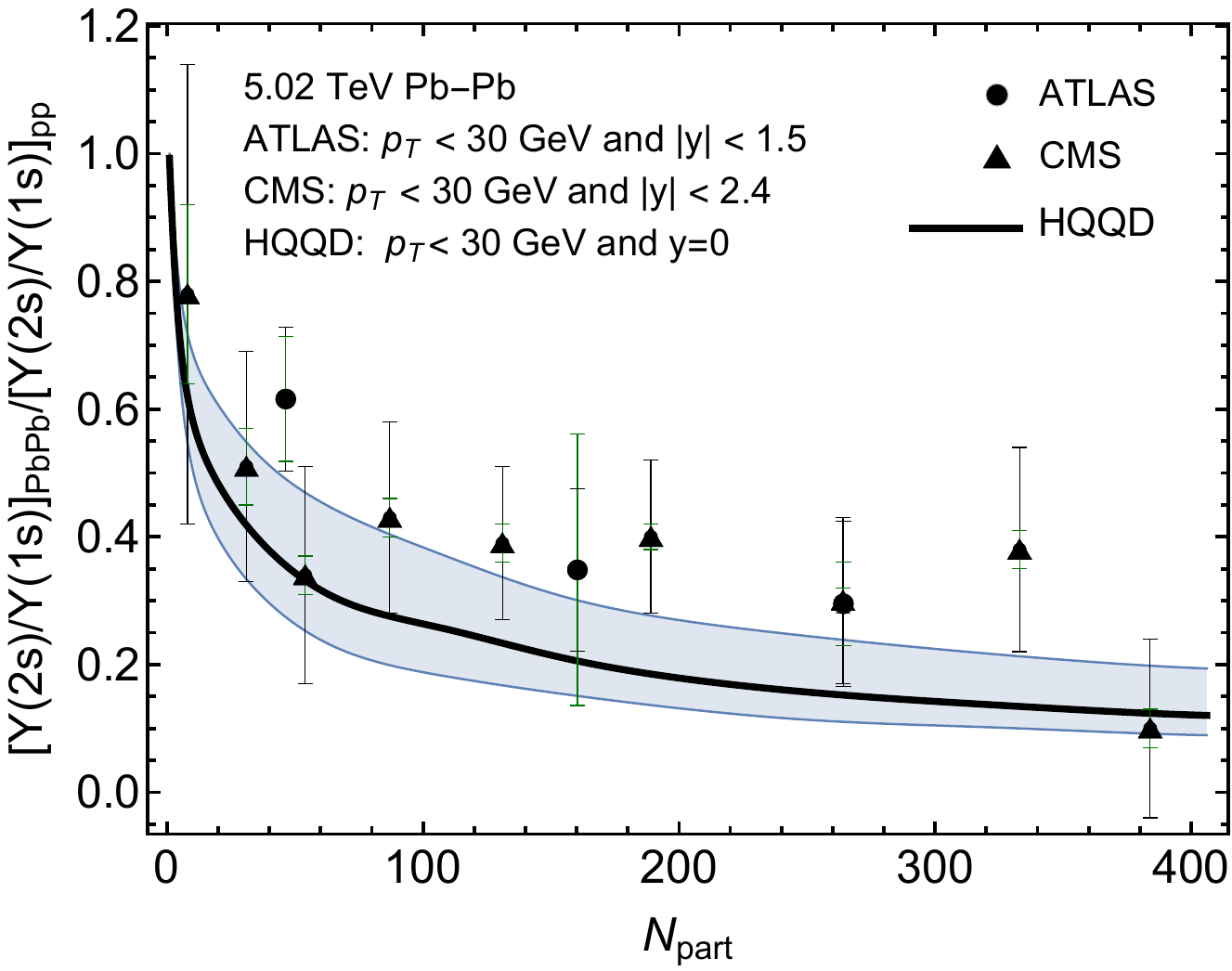} $\;\;\;\;$
		\includegraphics[width=0.48\linewidth]{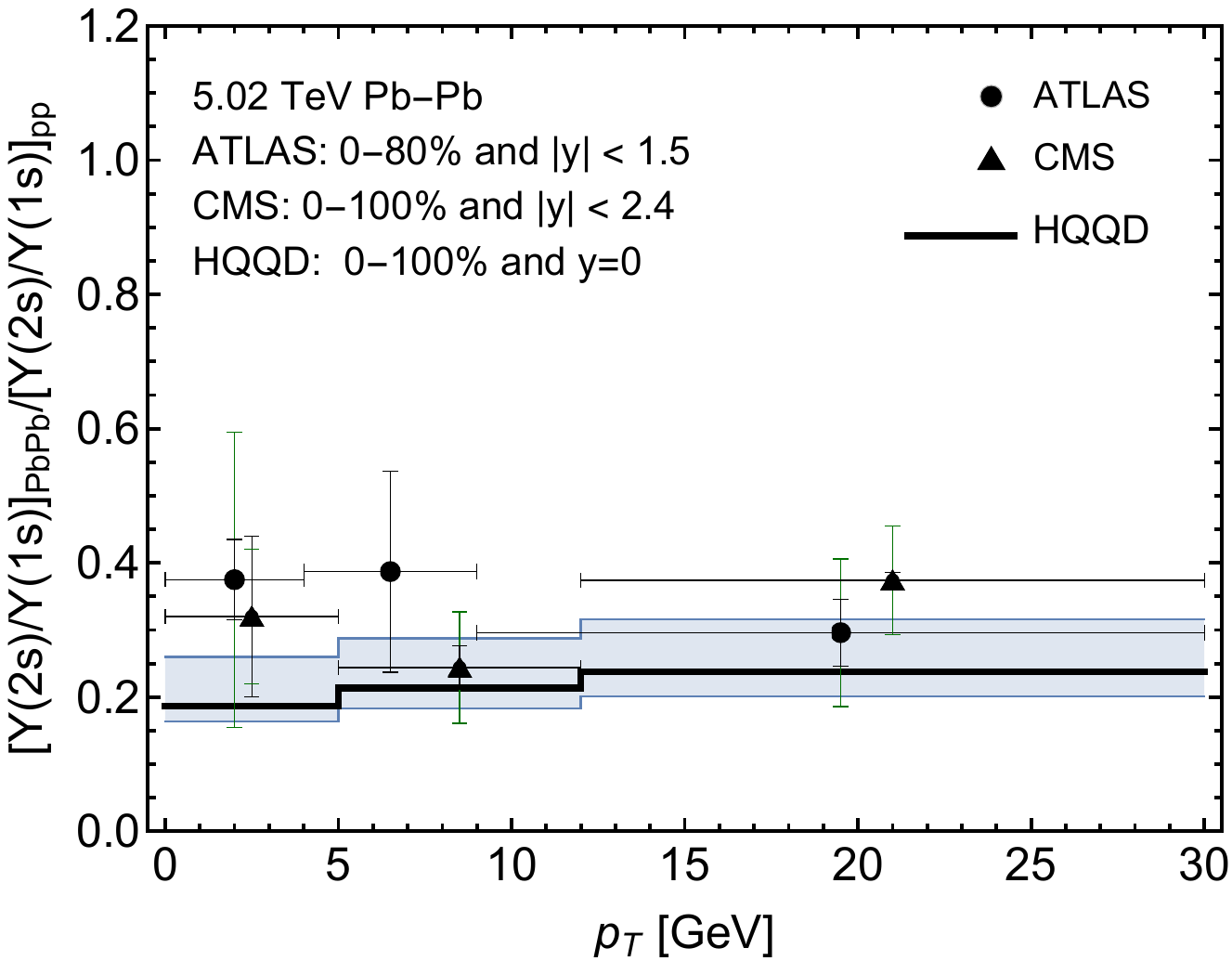}
	\end{center}
	\caption{Double ratio $[\Upsilon(2s)/\Upsilon(1s)]_\text{PbPb}/[\Upsilon(2s)/\Upsilon(1s)]_\text{pp}$ as a function of $N_\text{part}$ (left panel) and $p_T$ (right panel).  The solid line shows the prediction of HQQD.  The shaded bands indicate the variation of the HQQD result when changing the model parameters in the range specified in Eq.~\eqref{eq:bounds}.  The data points are from the ATLAS~\cite{ATLAS5TeV} and CMS ~\cite{Sirunyan:2017lzi} collaborations.  Systematic and statistical experimental uncertainties are indicated by the green and black error bars, respectively.}
	\label{fig:doubleratio} 
\end{figure}

Turning next to Fig.~\ref{fig:raavsnpartandpt} (right), we present the HQQD predictions for the transverse-momentum dependence of $R_{AA}[\Upsilon]$.  To obtain the HQQD predictions for this panel, we averaged over centrality with a weight function $w(c) = \exp(-c/20)$, with $c \in [0,100]$. This weight function allows us to mimic the experimentally observed distribution of the number of $\Upsilon$ states versus centrality~\cite{Chatrchyan:2012np}.  As can be seen from Fig.~\ref{fig:raavsnpartandpt} (right), HQQD predicts a very weak dependence of $R_{AA}[\Upsilon]$ on $p_T$, with only a small decrease in $R_{AA}[\Upsilon]$ at momentum less than the mass scale of the bottomonium states.  One can understand the increased suppression at low-$p_T$ in terms of the average effective lifetime of the quantum wave-packets in the QGP.  For low-momentum wavepackets, their effective lifetime in the QGP is longer due to their lower velocities and, hence, they experience stronger suppression than high-momentum wave-packets which can escape from the QGP more quickly.

In Figure \ref{fig:doubleratio} we plot the HQQD prediction for the $\Upsilon(2s)$ to $\Upsilon(1s)$ double ratio as a function of $N_\text{part}$(left panel) and $p_T$ (right panel). We compare these predictions with data reported by the ATLAS and CMS collaboration in Refs.~\cite{ATLAS5TeV,Sirunyan:2017lzi}.  The band shown in the Figure provides an estimate of our systematic uncertainty.  As can be seen from both of the panels of this Figure, we see further evidence that HQQD is predicting too much $\Upsilon(2s)$ suppression compared to the experimental data.   It is unclear at this moment time what is the source of this discrepancy.  We will return to this question in the conclusions where we discuss the possible impact of including stochastic in-medium color and angular momentum transitions in a more complete manner.

\begin{figure}[t]
	\begin{center}
		\includegraphics[width=0.45\linewidth]{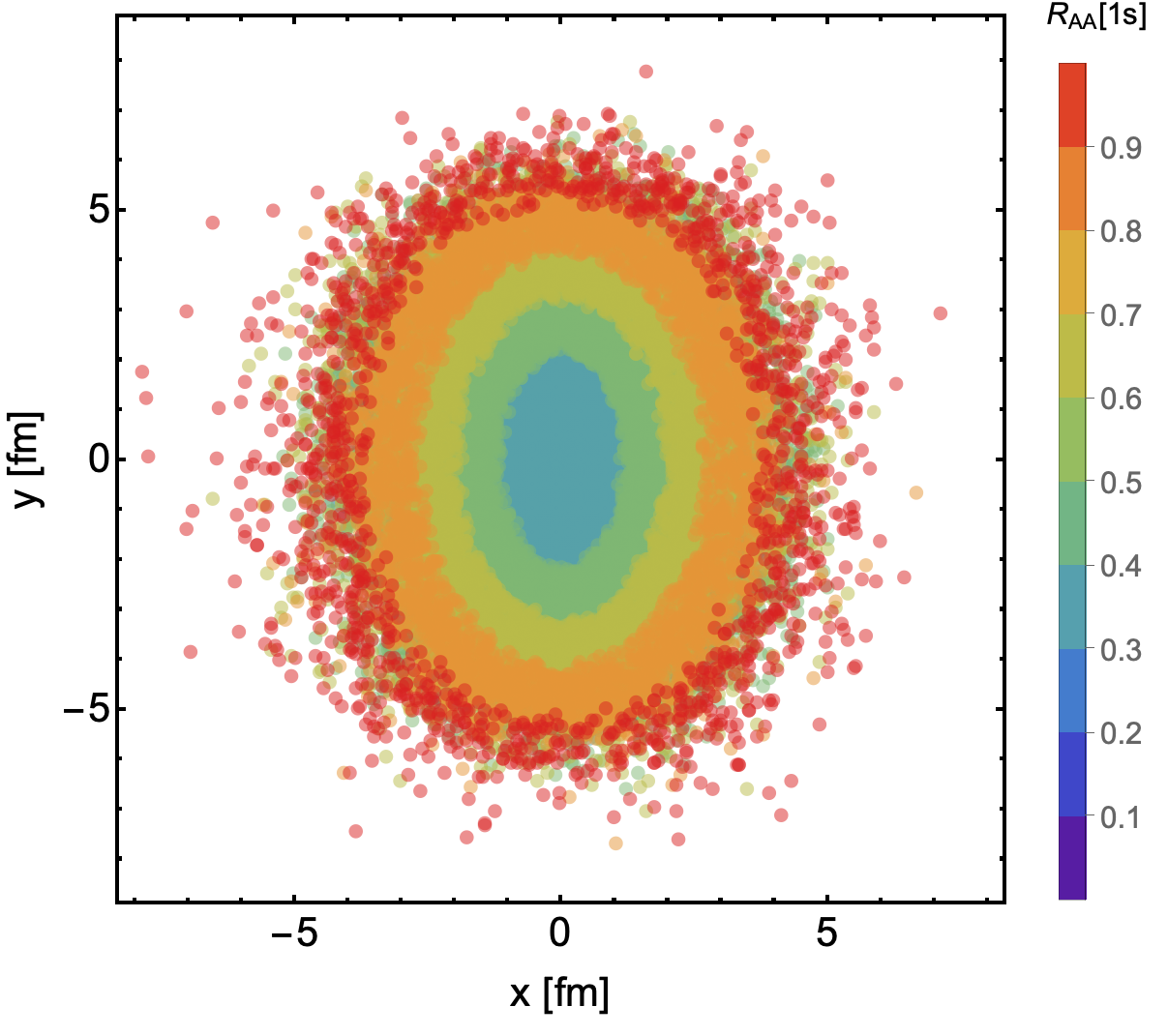} $\;\;\;\;$
		\includegraphics[width=0.45\linewidth]{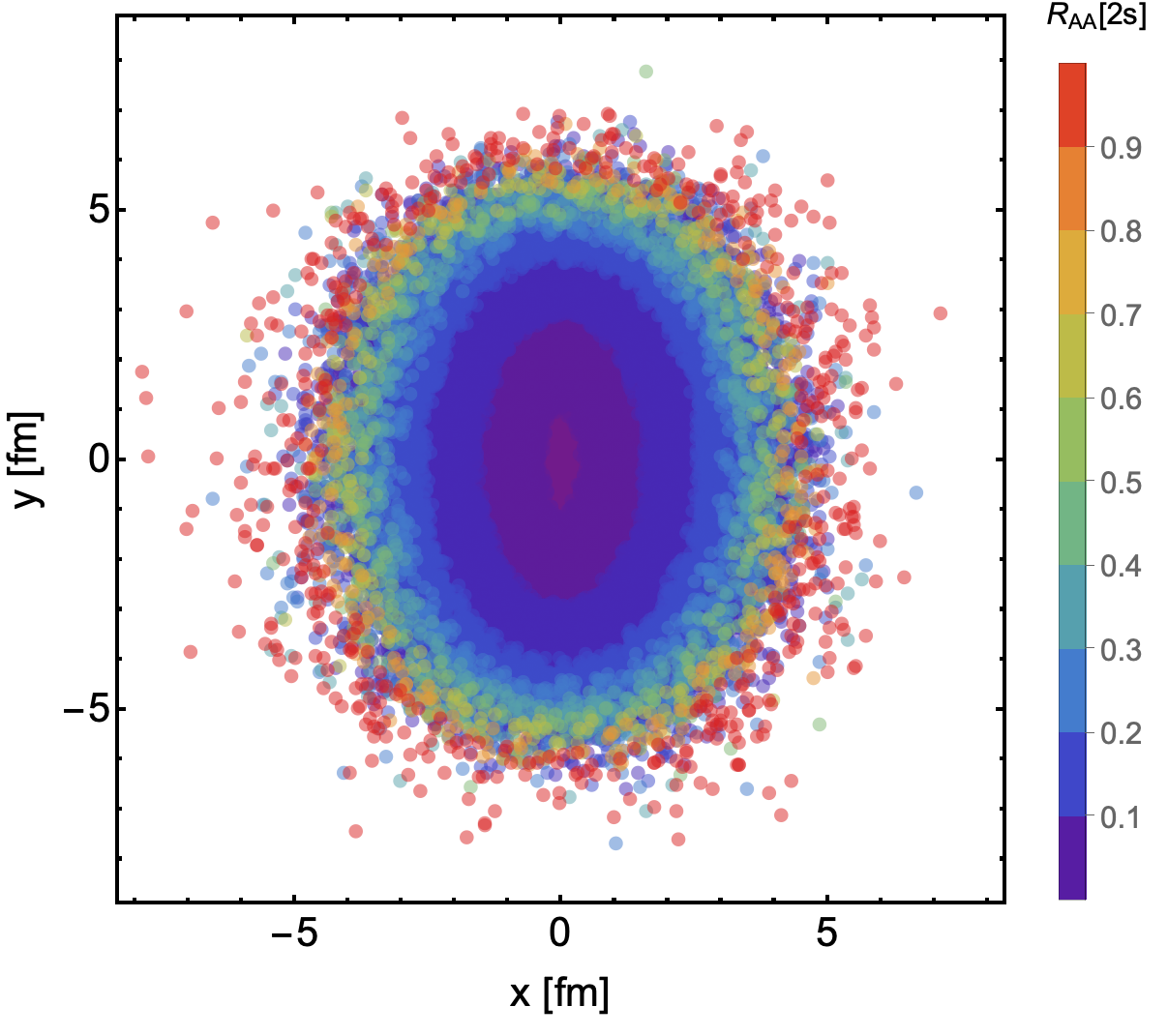}
	\end{center}
	\vspace{-5mm}
	\caption{Visualization of the starting positions of $\Upsilon(1s)$ (left) and $\Upsilon(2s)$ states (right).  In both cases we selected all states with $0 < p_T < 40$ GeV.  Color coding shows the $R_{AA}$ for each particle.} 
	\label{fig:tomo1}
\end{figure}

In order to gain further insight into the spatial-distribution of bottomonium suppression in the QGP one can construct ``tomographic'' plots showing the starting positions of the sampled bottomonium states, color-coded by the final, time-evolved $R_{AA}$ for that state.  From such plots one can learn where in the QGP bottomonium states are most strongly suppressed and where those with only weak-suppression emerge from.  A priori, one would expect trajectories that experience the strongest suppression to be those initially produced in the central of the fireball and trajectories with the weakest suppression to be those initially produced on the periphery of the collision.  This expectation is indeed observed in our simulations.  In Fig.~\ref{fig:tomo1} we show such a tomographic plot for the $\Upsilon(1s)$ (left) and the $\Upsilon(2s)$ (right).  For both panels in this figure, we considered a Pb-Pb collision with impact parameter of $b \simeq 9.2$ fm.  Comparing the left and right panels we firstly see the increased overall suppression of the $\Upsilon(2s)$ relative to the $\Upsilon(1s)$.  In both panels, we see that states with $R_{AA}$ close to one come from initial production near the surface, while those experiencing the strongest suppression come from the center of the fireball as expected.  

\begin{figure}[t]
	\begin{center}
		\includegraphics[width=0.95\linewidth]{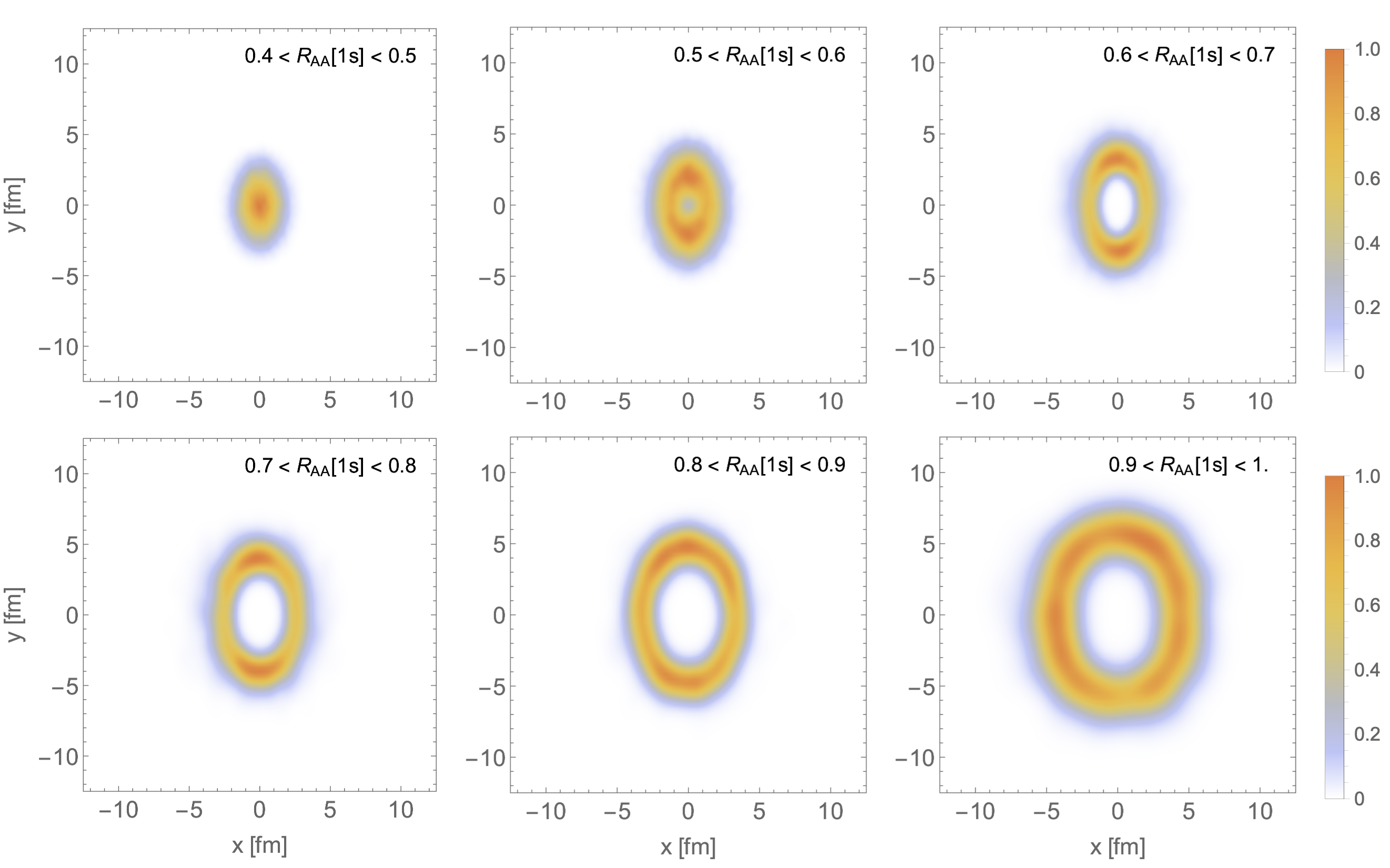}
	\end{center}
	\vspace{-5mm}
	\caption{Visualization of the initial probability distribution function for the starting positions of $\Upsilon(1s)$.  Panels show distribution for states experiencing a given $R_{AA}$ in 10\% intervals, e.g. $0.4 < R_{AA} < 0.5$.  For this visualization, we selected all states with $0 < p_T < 40$ GeV.  Coloring encodes the normalized probability in each $R_{AA}$ range.}
	\label{fig:tomo2}
\end{figure}

In order to better visualize the probability density for particles in each $R_{AA}$-class, in Fig.~\ref{fig:tomo2} we have plotted each $R_{AA}$-class in a separate panel with the panels corresponding to 0.1 intervals in $R_{AA}$ in the range $0.4 \leq R_{AA} \leq 1$.  In each panel we show the normalized probability distribution function (PDF) for each $R_{AA}$-class.  As in Fig.~\ref{fig:tomo1} we see that the suppression of the states is correlated with the initial geometry and that states with weak suppression typically come from the surface of the fireball.  Statistically, one can see, for example, that there is little overlap between the PDFs for particles with $R_{AA} \geq 0.7$ and $R_{AA} \leq 0.5$.  By comparing the production of particles in these two classes, one could investigate potential differences in bottomonium production in the central and peripheral regions of the QGP.  For now, however, such plots are mostly useful in making sure that our understanding of suppression dynamics is correct.

\begin{figure}[t]
	\begin{center}
		\includegraphics[width=\linewidth]{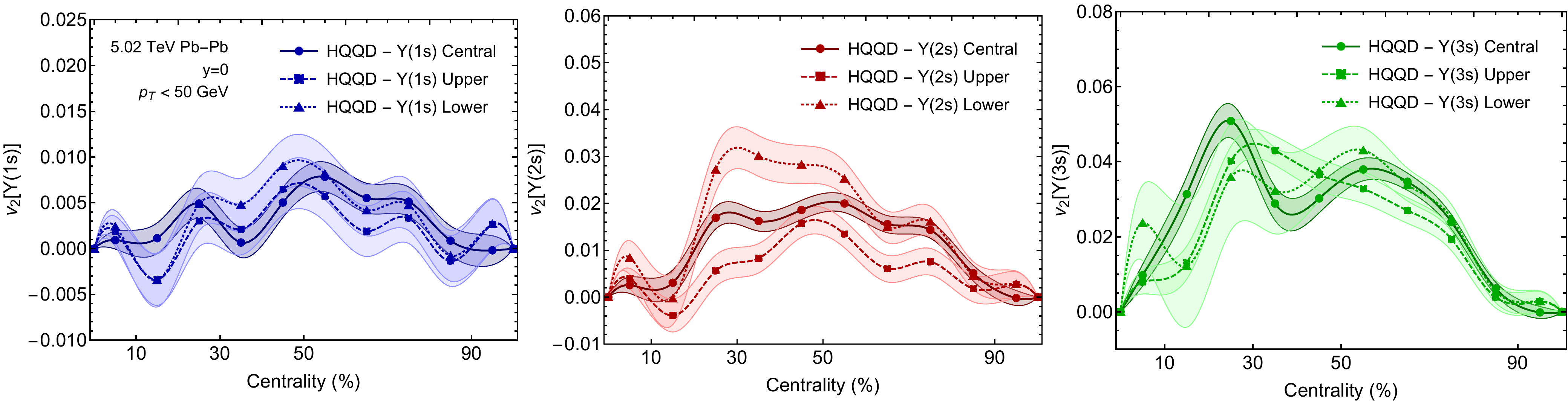}
	\end{center}
	\vspace{-5mm}
	\caption{Elliptic flow for $s$-wave bottomonium states as a function of centrality. Solid lines and bands show spline-interpolated results for the mean and statistical uncertainty of the mean obtained from HQQD.The upper (squares), central (circles), and lower (triangles) indicate the variation of the HQQD result when changing the model parameters in the range specified in Eq.~\eqref{eq:bounds}. Points show results obtained in equally spaced bins of 10\% centrality from 0-100\%.} 
	\label{fig:v2_swave}
\end{figure}

In Fig.~\ref{fig:v2_swave}, we present the HQQD predictions for centrality dependence of the elliptic flow of $\Upsilon(1s)$, $\Upsilon(2s)$, and $\Upsilon(3s)$ states. In this figure, we take $p_T < 50$ GeV and compute $v_2$ in 10 equally spaced centrality bins from 0-100\%.  The bands show the statistical uncertainty associated with the mean values extracted in each bin. The upper (squares), central (circles), and lower (triangles) lines correspond to the three different cases of varying the model parameters specified in Eq.~\eqref{eq:bounds}, respectively.  For each of the three choices of model parameters shown in Fig.~\ref{fig:v2_swave}, one sees an ordering of the elliptic flow of the states with the $\Upsilon(3s)$ state having the largest flow and the $\Upsilon(1s)$ the smallest.  This result is expected since excited states are more suppressed and hence experience a larger in-medium path-length dependence.  For each of the states shown in Fig.~\ref{fig:v2_swave} one notices that the oscillations in $v_2$ (to be discussed later) depend on the choice of the model parameters.  This can be attributed to the change in the effective path-length of the quantum wave-packets travel through the medium when changing $\tau_{\rm med}$.  Finally, we note that $v_2$ is more sensitive to these oscillations because it explicitly involves, for example, differences between the survival probability along the short and long sides of the QGP.  In $R_{AA}$ one sums over all angles and averages and, as a result, these oscillations are smoothed out in the sum over quantum wave-packet trajectories.  On the contrary, since for $v_2$ different angles contribute with different weights/signs, it is naturally more sensitive to small differences in the survival probability and hence is more sensitive to these oscillations.

\begin{figure}[t]
	\begin{center}
		\includegraphics[width=0.65\linewidth]{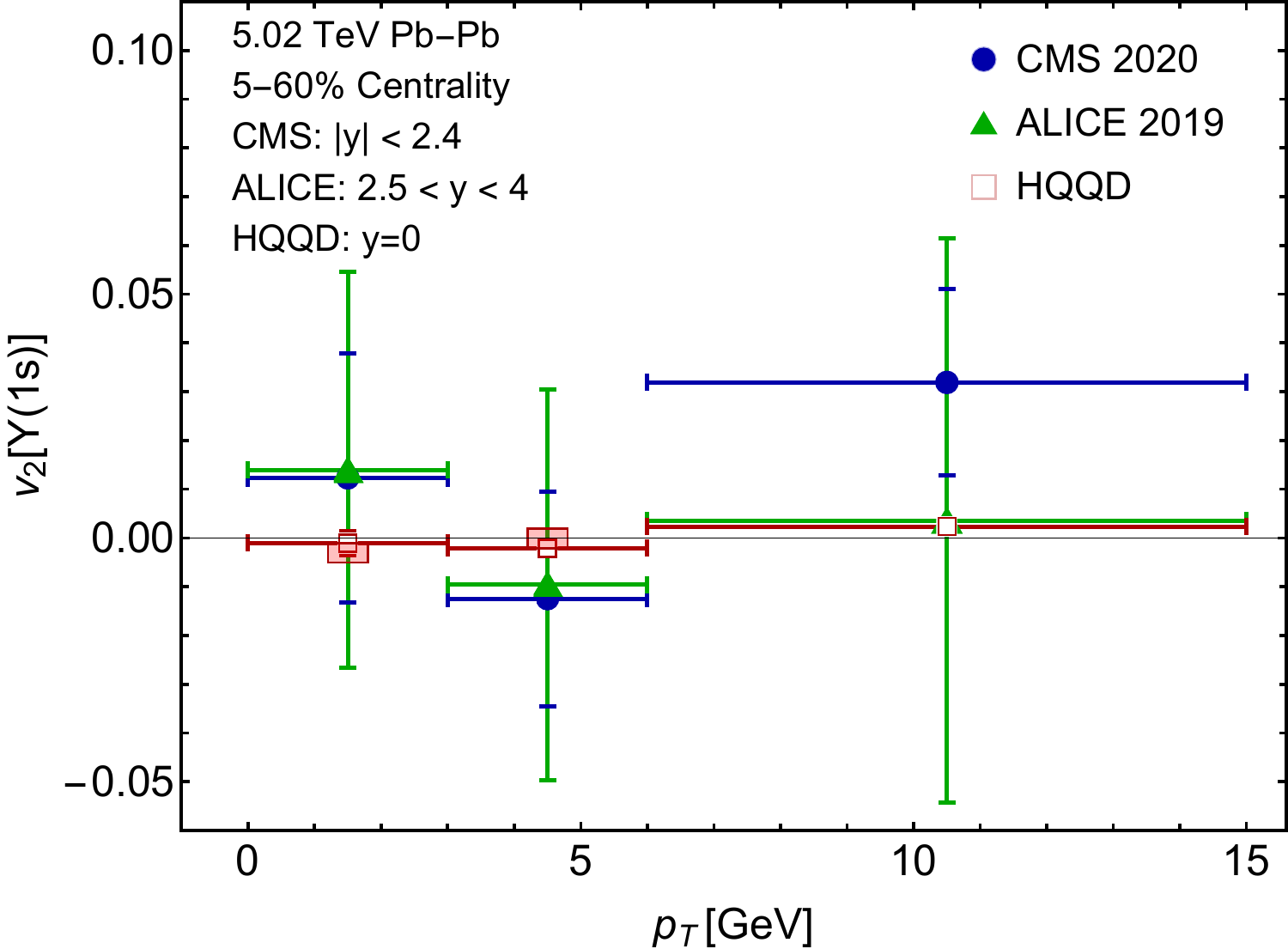}
	\end{center}
	\caption{The elliptic flow $v_2[\Upsilon(1s)]$ as a function of $p_T$ in three $p_T$-bins.  Open red squares are HQQD predictions and the data are from the ALICE \cite{Acharya:2019hlv} and CMS \cite{Sirunyan:2020qec} collaborations.  The shaded red boxes associated with each HQQD prediction indicate the systematic variation observed when changing the model parameters in the range specified in Eq.~\eqref{eq:bounds}.  Note that, for the highest $p_T$ bin, the systematic error estimate is so small that it is indistinguishable from the horizontal error bar.} 
	\label{fig:v21s}
\end{figure}

\begin{figure}[t!]
	\begin{center}
		\includegraphics[width=0.65\linewidth]{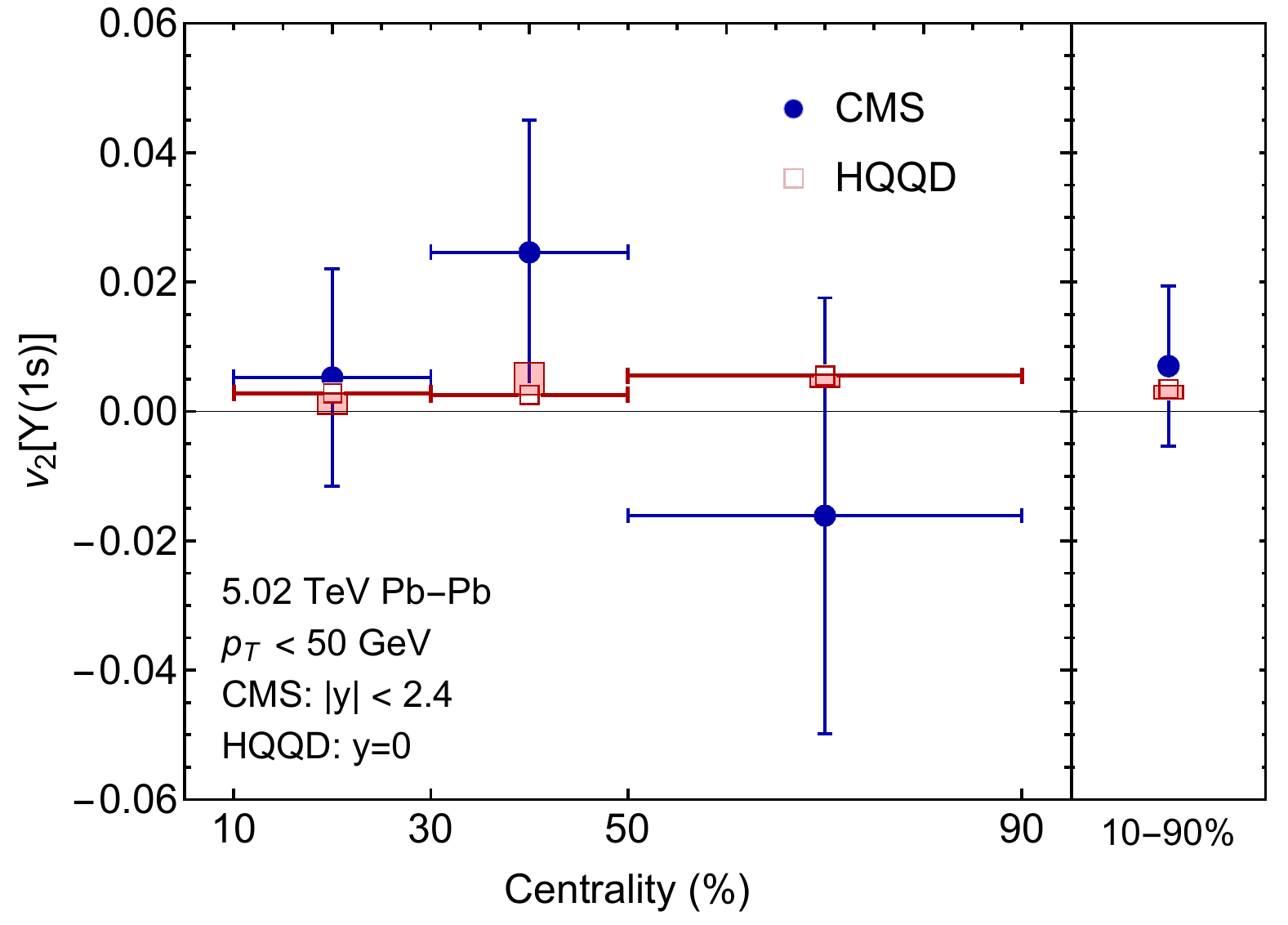}
	\end{center}
	\caption{Centrality dependence of $v_2[\Upsilon(1s)]$ shown in 10-30\%, 30-50\%, 50-90\%, and 10-90\% centrality bins. Open squares are predictions of HQQD.  The shaded red boxes associated with each HQQD prediction indicate the systematic variation observed when changing the model parameters in the range specified in Eq.~\eqref{eq:bounds}.  }
	\label{fig:v21sCent} 
\end{figure}

From Fig.~\ref{fig:v2_swave}, we also observe that the elliptic flow for all states goes to zero for central collisions (centrality = 0\%).  This is expected using the smooth (optical) Glauber model used herein for the hydrodynamic initial condition.  In the future it will be interesting to assess the degree of $\Upsilon$ elliptic flow in the most central classes taking into account also geometrical fluctuations in the initial condition (energy density, flow velocities, etc.).  We note, however, that since we use smooth hydrodynamic initial conditions $v_2$ must go to zero for central collisions.  This can be used as a check on our calculation of $v_2[\Upsilon]$ for all three states.  We also note that in the opposite limit, that of ultra-peripheral collisions (centrality = 100\%), one expects the $v_2$ for all states to go to zero since the QGP lifetime goes to zero in this limit.  This again provides a non-trivial test of our computation of $v_2$ for bottomonium states.

As alluded to above, from Fig.~\ref{fig:v2_swave} one notices that the dependence of $v_2$ on centrality can contain oscillations (non-monotonic behavior).  These oscillations arise in HQQD due to quantum oscillations in the overlaps of the in-medium bottomonium states with their corresponding vacuum states.  Wave-packets which experience different effective path-lengths can exit the QGP at different points during this quantum oscillation.  Since the characteristic period for these oscillations is on the order of 1 fm/c, this can have an observable effect on the HQQD predictions for $v_2$ as a function of centrality.

\begin{figure}[t]
	\begin{center}
		\includegraphics[width=0.65\linewidth]{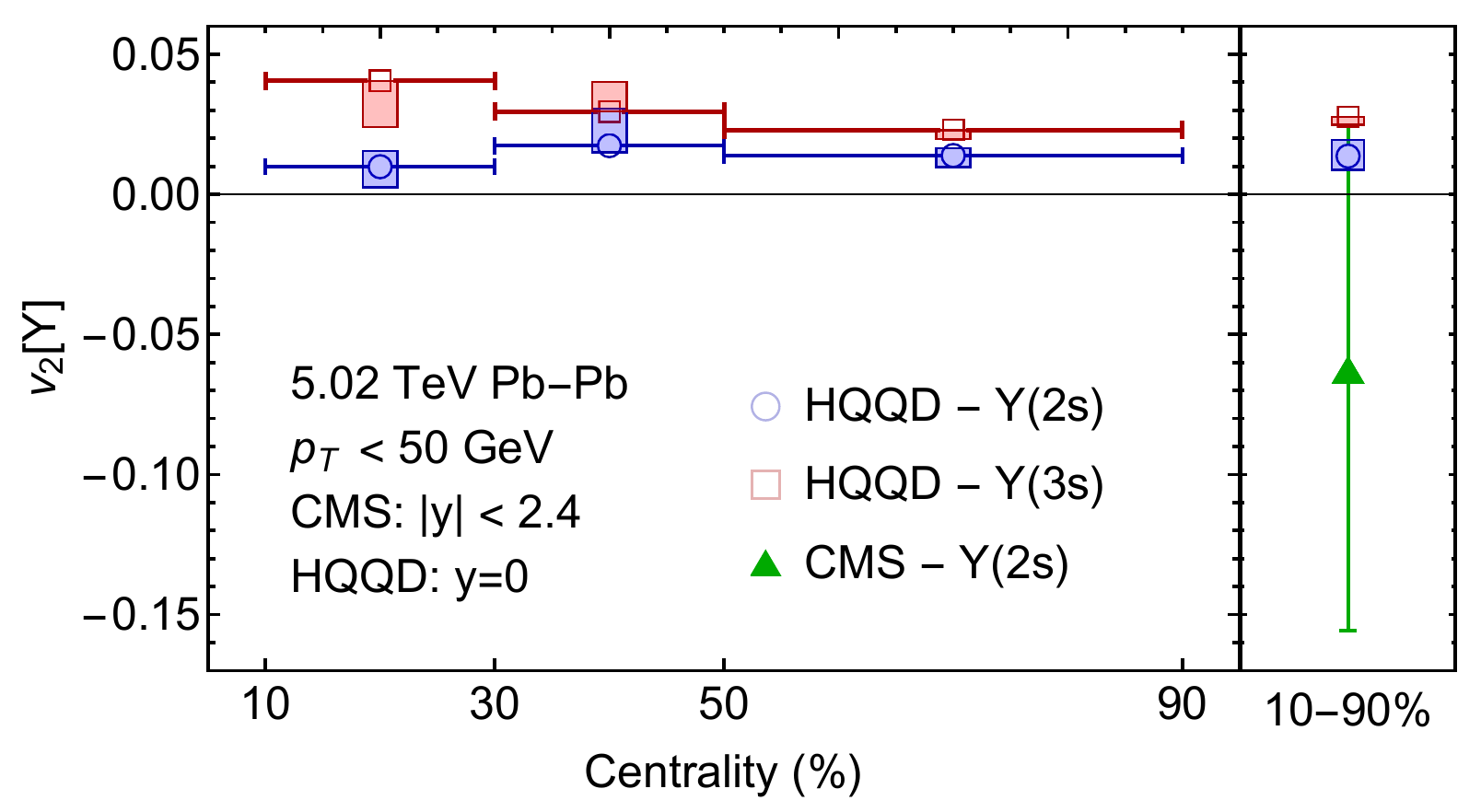}
	\end{center}
	\caption{Centrality dependence of $v_2[\Upsilon(2s)]$ and $v_2[\Upsilon(3s)]$ in the same centrality bins as Fig.~\ref{fig:v21sCent}.  Open symbols are predictions of HQQD.  In the 10-90\% class we include recent data reported by the CMS collaboration for integrated $v_2[\Upsilon(2s)]$ \cite{Sirunyan:2020qec}.  The shaded red and blue boxes associated with each HQQD prediction indicate the systematic variation observed when changing the model parameters in the range specified in Eq.~\eqref{eq:bounds}. } 
	\label{fig:v2excitedCent} 
\end{figure}

A comparison of HQQD predictions for $v_2[\Upsilon(1s)]$ with experimental data collected by the ALICE \cite{Acharya:2019hlv} and CMS \cite{Sirunyan:2020qec} collaborations in three different transverse momentum bins, 0-4, 4-6, and 6-15 GeV, is shown in Fig.~\ref{fig:v21s}. The HQQD predictions and experimental results are integrated over centrality in the range \mbox{5-60\%}. One sees from this figure that HQQD predicts results consistent with $v_2$ being zero, slightly negative, and small but positive value in the lowest $p_T$ bin, central $p_T$ bin, and  highest $p_T$ bin, respectively. A similar trend has also been predicted earlier by an adiabatic approximation based model in Ref.~\cite{Bhaduri:2020lur}.   According to Ref.~\cite{Bhaduri:2020lur}, the negative $v_2$ observed in the central $p_T$ bin is related to the transverse expansion of the QGP overtaking bottomonia states which have escaped from near the surface of the QGP.   To demonstrate that this was indeed the cause for the negative $v_2$, the authors Ref.~\cite{Bhaduri:2020lur} considered the case in which the transverse expansion of the QGP was turned off, finding that in this case $v_2$ was always positive.   Returning to the theory to data comparison, Fig.~\ref{fig:v21s} demonstrates that HQQD has a reasonable agreement with the available experimental data given current experimental uncertainties.   Finally, we note that even given the experimental uncertainties, we see a similar trend in the experimental results as a function of $p_T$ as predicted by HQQD.

\begin{figure}[t]
	\begin{center}
		\includegraphics[width=0.65\linewidth]{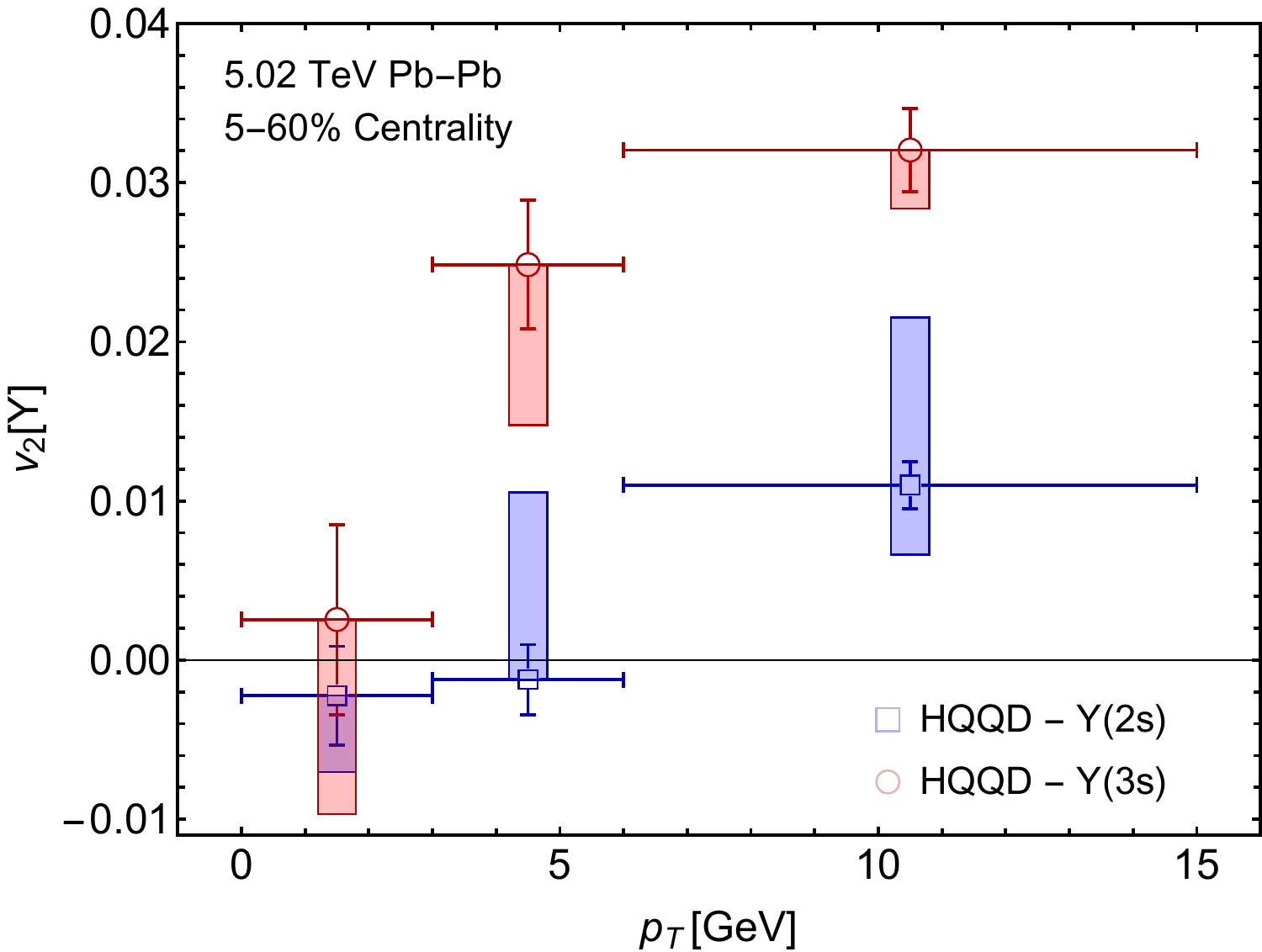}
	\end{center}
	\caption{HQQD predictions for the elliptic flow of $\Upsilon(2s)$ and $\Upsilon(3s)$ states in the 5-60\% centrality bin.  The transverse momentum bins are the same as used in Fig.~\ref{fig:v21s}.  The shaded red and blue boxes associated with each HQQD prediction indicate the systematic variation observed when changing the model parameters in the range specified in Eq.~\eqref{eq:bounds}. }
	\label{fig:v22s3s}
\end{figure}

In Fig.~\ref{fig:v21sCent}, we present the centrality dependence of $v_2[\Upsilon(1s)]$ in 10-30\%, 30-50\%, 50-90\%, and 10-90\% centrality bins and compare our HQQD predictions with experimental data from the CMS collaboration~\cite{Sirunyan:2020qec}. We show a comparison of the HQQD prediction with the experimental result integrated over 10-90\% centrality in the right panel. As can be seen from this figure, in the case of the integrated $v_2[\Upsilon(1s)]$ in the 10-90\% bin (right panel), there is quite reasonable agreement, within experimental uncertainties, between the HQQD prediction and the data reported by the CMS collaboration. In the left panel of the Fig.~\ref{fig:v21sCent}, good agreement between HQQD predictions and CMS data can be seen in the 10-30\% bin, however, there are larger differences in the other two centrality bins but still within $2\sigma$ of the experimental data. Hopefully, higher experimental statistics from future runs will reduce the experimental uncertainties in the near future.

\begin{table}[t!]
\begin{center}
\def\arraystretch{1.4}
{\scriptsize
\begin{tabular}{|c|c|c|c|}
\hline
{\bf ~Observable~} & {\bf ~Source/Cuts~}  &{\bf ~Experiment/HQQD~}  \\
\hline
$R_{AA}[\Upsilon(1s)]$ & ALICE 0-90\%  \cite{Acharya:2018mni}& 0.37 $\pm$ 0.02 $\pm$ 0.03  \\
& $p_T < 15$ GeV & 0.301 $\pm$ 0.0002 $\pm\,^{0.09}_{0.12}$.  \\
\hline
$R_{AA}[\Upsilon(1s)]$ & ATLAS 0-80\% \cite{ATLAS5TeV} & 0.32 $\pm$ 0.02 $\pm$ 0.05   \\
& $p_T < 30$ GeV & 0.3046 $\pm$ 0.0001 $\pm\,^{0.08}_{0.12}$ \\
\hline
$R_{AA}[\Upsilon(1s)]$ & CMS 0-100\% \cite{Sirunyan:2018nsz}& 0.376 $\pm$ 0.013 $\pm$ 0.035  \\
& $p_T < 30$ GeV & 0.312 $\pm$ 0.0002 $\pm\,^{0.085}_{0.12}$ \\
\hline
$R_{AA}[\Upsilon(2s)]$ & ALICE 0-90\% \cite{Acharya:2018mni}& 0.10 $\pm$ 0.04 $\pm$ 0.02  \\
& $p_T < 15$ GeV & 0.05923 $\pm$ 0.00005 $\pm\,^{0.059}_{0.026}$ \\
\hline
$R_{AA}[\Upsilon(2s)]$ & ATLAS 0-80\% \cite{ATLAS5TeV} & 0.11 $\pm$ 0.04 $\pm$ 0.04   \\
& $p_T < 30$ GeV & 0.07920 $\pm$ 0.00006 $\pm\,^{0.058}_{0.039}$ \\
\hline
$R_{AA}[\Upsilon(2s)]$ & CMS 0-100\% \cite{Sirunyan:2018nsz}& 0.117 $\pm$ 0.022 $\pm$ 0.019  \\
& $p_T < 30$ GeV & 0.0679 $\pm$ 0.00004 $\pm\,^{0.059}_{0.026}$ \\
\hline
$R_{AA}[\Upsilon(3s)]$ & CMS 0-100\% \cite{Sirunyan:2018nsz}& 0.022 $\pm$ 0.038 $\pm$ 0.016   \\
& $p_T < 30$ GeV & 0.03622 $\pm$ 0.00004 $\pm\,^{0.032}_{0.012}$ \\
\hline
$\underline{[\Upsilon(2s)/\Upsilon(1s)]_\text{PbPb}}$ & ATLAS 0-80\% \cite{ATLAS5TeV}& 0.35 $\pm$ 0.12 $\pm$ 0.12   \\
$[\Upsilon(2s)/\Upsilon(1s)]_\text{pp}$ & $p_T < 30$ GeV & 0.193 $\pm$ 0.0002 $\pm\,^{0.08}_{0.04}$ \\
\hline
$\underline{[\Upsilon(2s)/\Upsilon(1s)]_\text{PbPb}}$ & CMS 0-100\% \cite{Sirunyan:2017lzi}& 0.308 $\pm$ 0.055 $\pm$ 0.001  \\
$[\Upsilon(2s)/\Upsilon(1s)]_\text{pp}$ & $p_T < 30$ GeV & 0.2175 $\pm$ 0.0002 $\pm\,^{0.075}_{0.031}$ \\
\hline
$v_2[\Upsilon(1s)]$ & ALICE 5-60\% \cite{Acharya:2019hlv}& -0.003 $\pm$ 0.030 $\pm$  0.006  \\
& $2 < p_T < 15$ GeV & 0.0009 $\pm$ 0.0009 $\pm\,^{0.001}_{0.00005}$ \\
\hline
$v_2[\Upsilon(1s)]$ & CMS 10-90\% \cite{Sirunyan:2020qec}& 0.007 $\pm$ 0.011 $\pm$ 0.005 \\
 & $p_T < 30$ GeV & 0.003 $\pm$ 0.0007 $\pm\,^{0.0006}_{0.0013}$  \\
 \hline
$v_2[\Upsilon(2s)]$ & CMS 10-90\% \cite{Sirunyan:2020qec}& -0.063 $\pm$ 0.085 $\pm$ 0.037  \\
 & $p_T < 30$ GeV & 0.0127 $\pm$  0.0008 $\pm\,^{0.006}_{0.005}$ \\
 \hline
 $v_2[\Upsilon(3s)]$ & HQQD 10-90\% & N/A \\
 & $p_T < 30$ GeV & 0.0264 $\pm$ 0.0011 $\pm\,^{0.0006}_{0.003}$ \\
 \hline
\end{tabular}
}
\end{center}
\caption{Comparison of HQQD predictions for integrated $R_{AA}[\Upsilon]$ and $v_2[\Upsilon]$ with available experimental data.  In each row, the first column indicates the observable, the second column indicates the source of the experimental result and relevant cuts, and the third column shows the experimental result on the first line and the HQQD prediction on the second line. For both HQQD predictions and the experimental results, the first uncertainty reported is statistical uncertainty and the second is systematic uncertainty.  In the case of HQQD, the systematic uncertainities were estimated by varying the HQQD model parameters in the range specified in Eq.~\eqref{eq:bounds}.}
\label{tab:comp}
\end{table}

In Fig.~\ref{fig:v2excitedCent}, we show the centrality dependence of $v_2[\Upsilon(2s)]$ and $v_2[\Upsilon(3s)]$ in the same centrality bins as used in Fig.~\ref{fig:v21sCent}. One centrality-integrated data point for $v_2[\Upsilon(2s)]$ is currently available from the CMS collaboration and has been shown as a green triangle in the 10-90\% panel (right). As can be seen from the right panel of the figure, the integrated result for $v_2[\Upsilon(2s)]$ is within the experimental uncertainties, however, at the very top end of them. We again hope that more statistics will help in making more constraining comparisons in the near future.  Note that for  $v_2[\Upsilon(2s)]$ and $v_2[\Upsilon(3s)]$ one sees a larger systematic uncertainty than for $v_2[\Upsilon(1s)]$.

In Fig.~\ref{fig:v22s3s}, we present HQQD predictions for the elliptic flow of $\Upsilon(2s)$ and $\Upsilon(3s)$ states as a function of transverse momentum in the 5-60\% centrality bin. As can be seen from this figure, a sizable $v_2$ is predicted by HQQD for the $\Upsilon(3s)$ because of the strong path length dependence of $R_{AA}[\Upsilon(3s)]$ between the short and long sides of the QGP fireball. Similar to the $\Upsilon(1s)$, for the central values of our model parameters, HQQD predicts a negative $v_2$ for the $\Upsilon(2s)$ in the lowest two $p_T$-bins. The reason behind this negative $v_2$ is again that the transverse expansion of the QGP occurs more rapidly along the short side than the long side which has the affect of overtaking bottomonium states which had previously escaped the QGP with $\phi \sim 0$. One can see that HQQD makes a prediction that $v_2$ is positive in the highest $p_T$-bin for both states.  Turning to the model variation shown as red and blue box in Fig.~\ref{fig:v22s3s}, one sees that in the 4-6 GeV $p_T$ bin the variation of the model parameters can result in a positive $v_2[\Upsilon(2s)]$.  This large systematic HQQD uncertainty is related to the quantum-mechanical oscillations discussed previously in the context of Fig.~\ref{fig:v2_swave}.  Given these larger uncertainties, however, one still observes an ordering of the elliptic flow of the excited states with $v_2[\Upsilon(3s)] > v_2[\Upsilon(2s)]$ in the highest two $p_T$ bins.  

Finally, in Table \ref{tab:comp} we compare the predictions of HQQD for various observables with the corresponding experimental results from the ALICE, ATLAS, and CMS collaborations. In each row of the table, the results are integrated over centrality and transverse momentum in the ranges shown in the middle column, the first line of the third column is the experimental data (if available), and the second line of the third column shows the HQQD prediction. The rapidity cuts are $2.5 < y < 4.0$, $|y| < 1.5$, and $|y| < 2.4$ for the ALICE, ATLAS, and CMS collaborations, respectively. As can be seen from this table, all HQQD predictions are within the combined experimental statistical and systematic uncertainties. However, as can be seen most clearly from our comparisons of the integrated $R_{AA}[\Upsilon(2s)]$ and $\Upsilon(2s)$ to $\Upsilon(1s)$ double ratio, HQQD predicts too much $\Upsilon(2s)$ compared to the current experimental data.

\section{Conclusions and outlook}
\label{sec:conclusions}

In this paper, we have extended our previous work \cite{Islam:2020gdv} by allowing for variation in the underlying HQQD model parameters in order to assess the impact such variation has on HQQD predictions for $R_{AA}[\Upsilon]$ and $v_2[\Upsilon]$.  In addition, herein, we provided more details concerning the numerical method used in HQQD and the method by which final state feed down effects were included.  Similar to our previous paper we used real-time quantum evolution averaged over a large set of bottomonium trajectories (2 million).  After averaging over all wave-packet trajectories, for each set of model parameters we were able to obtain precise estimates for $R_{AA}$.  For the central values of the model parameters, we found quite reasonable agreement with available experimental data for $R_{AA}[\Upsilon(1s)]$, however, we found that HQQD over-predicts the suppression of  $\Upsilon(2s)$ compared to experimental observations.  For the case in which we decreased the initialization time and increased the effective Debye mass, we found that the model predicts perhaps too much $\Upsilon(1s)$ suppression as can be seen from the lower limits indicated in Fig.~\ref{fig:raavsnpartandpt}.  For the opposite case in which we increased the initialization time and decreased the effective Debye mass, we found that the model under-predicts the amount of $\Upsilon(1s)$ suppression, while providing better agreement with data available for $R_{AA}[\Upsilon(2s)]$.  

Using this same set of model parameters, we then looked at the impact of our model assumptions on $v_2[\Upsilon]$.  We found that the results for $v_2[\Upsilon(1s)]$ were relatively insensitive to the choice of model parameters, with the maximum  $v_2[\Upsilon(1s)]$ as a function of centrality being on the order of 1\% in all three cases considered.  We found stronger variations in the HQQD predictions for $v_2[\Upsilon(2s)]$ and $v_2[\Upsilon(3s)]$, however, we still observed a maximal $v_2[\Upsilon(2s)]$ on the order of 3\% and $v_2[\Upsilon(2s)]$ on the order of 5\%.   For each of the sets, we found that the ordering $v_2[\Upsilon(1s)] \lesssim v_2[\Upsilon(2s)] \lesssim v_2[\Upsilon(3s)]$ remains valid.  This ordering is expected due to the fact that excited states experience more suppression and, hence, more path-length dependence of their propagation than the ground state.  All of our HQQD predictions for integrated suppression and elliptic flow of bottomonium states were finally summarized in Tab.~\ref{tab:comp} which includes estimates of both the systematic and statistical model uncertainties.

With respect to the HQQD over prediction for $\Upsilon(2s)$ suppression, we note that HQQD includes in-medium thermal noise only through the imaginary part of the potential included in the real-time dynamics.  This allowed us to extract the survival probability associated with the thermal-noise averaged quantum wave-function, however, does not fully take into account in-medium transitions between different singlet/octet and angular momentum states. There is currently ongoing work to include true stochastic noise at the level of the real-time Schr\"odinger equation solution, either through explicit noisy potentials \cite{Akamatsu:2011se,Akamatsu:2012vt,Rothkopf:2013ria,Rothkopf:2013kya,Kajimoto:2017rel,Akamatsu:2018xim,Miura:2019ssi,Sharma:2019xum} or through solution of the Lindblad equation for the reduced density matrix \cite{Akamatsu:2014qsa,Brambilla:2016wgg,Brambilla:2017zei,Blaizot:2017ypk,Blaizot:2018oev}.  Preliminary results obtained using the quantum trajectories method to solve the Lindblad equation indicate that more accurate inclusion of noise effects including, e.g., in-medium singlet-octet transitions, will result in only small changes in $R_{AA}[\Upsilon(1s)]$~\cite{Brambilla:2020qwo}.  Conversely, these studies indicate that the excited states can be less suppressed when including in-medium stochastic singlet-octet transitions.  This is expected due to the possibility of ``quantum regeneration'' in which a state transitions to a finite $\ell$ octet state and then back to a singlet $\ell=0$ state, either through single or multiple in-medium quantum state transitions~\cite{Brambilla:2020qwo}. It is possible that inclusion of the effect of in-medium quantum jumps will help to reduce the tension between HQQD predictions for $R_{AA}[\Upsilon(2s)]$ and current experimental data.

In closing, we note that herein we made use of a potential whose real part was obtained from the internal energy.  There were three reasons for choosing the internal energy over the free energy:  (1) it was found in prior studies using the adiabatic approximation that using the free energy resulted in too much suppression for all states (see Ref.~\cite{Strickland:2011aa}), (2) there is some theoretical support for this choice when the system is in equilibrium due to entropy considerations, see e.g. Refs.~\cite{PhysRevD.70.054507,PhysRevC.70.021901,SHURYAK200564}, and (3) this same potential has been used in many prior works that utilized the adiabatic approximation.  Another constraint on the potential was that it possesses the correct low temperature limit, i.e. that it should reduce to a real-valued Cornell potential tuned to the known spectrum of bottomonium bound states.  This was done because we wanted to (a) have a more realistic phenomenological description of bottomonium excited states and (b) match smoothly onto the known vacuum eigenstates which are used, in practice, to project out the final survival probabilities.  Of course, it would also be nice to add some non-perturbative contributions to the imaginary part of the potential, however, on this front things are much less constrained.  One possibility would be to use lattice extractions of the imaginary part of the potential similar to what was done in Ref.~\cite{Krouppa:2017jlg}.  Such investigations are planned for the future.
An additional possibility is that the choice of the model potential used herein can be improved to include, for example, the effects of non-equilibrium momentum-space anisotropies in the QGP.  The inclusion of momentum-space anisotropy has been shown to reduce quarkonium suppression for all states due to a reduction in the effective in-medium Debye mass \cite{Margotta:2011ta}.  We, again, leave this for future work.

\section*{Acknowledgments}

We thank the participants of the EMMI Rapid Reaction Task Force meeting on ``Suppression and (re)generation of quarkonium in heavy-ion collisions at the LHC'' for useful discussions.  We also thank the Ohio Supercomputer Center under the auspices of Project No.~PGS0253.  M.S. and A.I.  were supported by the U.S. Department of Energy, Office of Science, Office of Nuclear Physics Award No.~DE-SC0013470.

\bibliographystyle{JHEP}
\bibliography{hqqdJHEP}

\end{document}